\documentclass{aa} 

\usepackage{graphicx}
\usepackage{natbib}
\usepackage{txfonts}

\begin{document}

\title{Be stars: one ring to rule them all ?}

   \author{A.~Meilland \inst{1} 
            \and
          Ph.~Stee \inst{1} 
           \and
           J.~Zorec \inst{2}
            \and
          S.~Kanaan \inst{1}
          }

   \offprints{A. Meilland}

\institute{Observatoire de la C\^{o}te d'Azur-CNRS-UMR 6203 Avenue Copernic,
Grasse, France
 \and
 Institut d'Astrophysique de Paris, UMR7095 CNRS, Univ. P. \& M. Curie, Paris, France.}

   \date{Received; accepted }

   \abstract{}{We report the theoretical spectral energy distributions (SEDs), Br$\gamma$ line profiles, and visibilities for two scenarios 
that can explain the disk dissipation of active hot stars and account for the transition from the Be to the B spectroscopic phase.}
{We use the SIMECA code to investigate two scenarios, the first one where the disk is formed 
by successive outbursts of the central star. A low-density region is developing above the star and slowly grows outward and
forms a ring-like structure that will gradually excavate the disk. The second one has a slowly decreasing mass loss due for instance, to a decrease in the radiative force through an opacity change at the base of the photosphere, and may also be responsible 
for the vanishing of the circumstellar disk.} {We find that a clear signature of the disk dissipation following the ring scenario will be 
the disappearance of the high velocity tails in the emission lines and a nearly constant peak separation. Moreover, we found that,
following the ring-like scenario, the visibilities must show an increasing second lobe, an increase in the value of the first zero, and 
assuming an unresolved central star, a first zero of the visibility curves that appends at shorter baselines as far as the disk has been excavated. 
We propose to use the AMBER instrument on the VLTI to probe whether the ring scenario is the one to rule the Be phenomenon.}{}


   \maketitle
%

\section{Introduction}
According to advances in the study of stellar mass-loss phenomena since the discovery
of Be stars, different kinds of models have attempted to describe their 
circumstellar envelope (CE) or disk formation. Although they coexist, we may 
attempt to divide the mass-ejection phenomena into two broad classes: 1) 
continuing mass ejections and 2) discrete mass ejections. Even though one or the 
other of these mass-loss events might be prevalent, somewhat different types of CE 
structures could be envisioned:\par 
\begin{figure}[t]
  \begin{center}
      \includegraphics[height=8.cm]{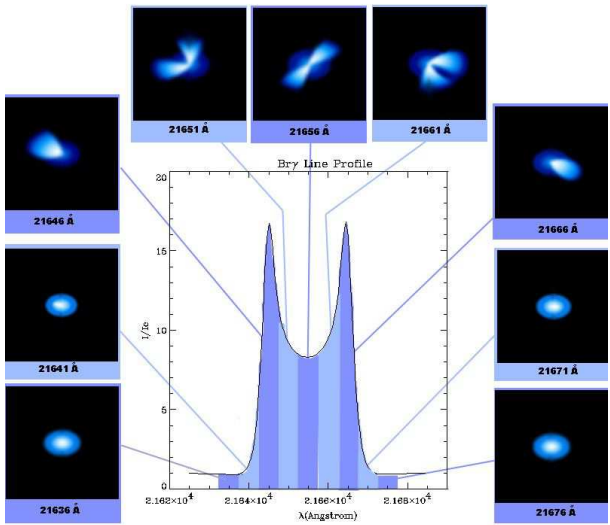}
  \end{center}
  \caption{Isovelocity intensity maps computed with the SIMECA code, seen with an inclination angle of 45$\degr$, and showing the effect of the disk kinematics (mainly Keplerian rotation) as a function of wavelength, 21636, 21641, 21646, 21651, 21656, 21661, 21666, 21671, and 21676 $\AA$ across the Br$\gamma$ line within a spectral bandwidth of 5 $\AA$.  }
\label{visi_baseline_radius}
\end{figure}

1) \textsc{Continuing mass ejections}. The most frequently suggested
mechanisms of CE formation evoke continuing mass ejections. Struve (1931, 1942)
imagined Be stars as secularly stable critical rotators that build a rotating extended envelope flattened towards the 
equator through equatorial ejection of mass(Limber (1964, 1967, 1969). Somewhat modified by the inclusion of an 
expansion velocity field, these CE structures were used by Marlborough (1969)
and Poeckert \& Marlborough (1987a, b) to predict observable spectroscopic, 
photometric, and polarimetric quantities. Thenceforth, several ad hoc flaring 
and flattened disc-shaped empirical CE models were explored in the literature
to see if the CEs were outwardly accelerated, decelerated, and/or rotationally 
supported (Waters 1986; Waters et al. 1987; Kastner \& Mazzali 1989; Hummel \& 
Dachs 1992; Hummel 1994; Hanuschik 1995, 1996; Wood et al. 1997). Thomas \& 
Doazan (1982) proposed an active spherical exophotospheric structural pattern, 
where the low-density outwardly accelerated hot chromospheric-coronal zone is 
followed by a cool and decelerated high density region. However, 
interferometric and polarimetric observations, as well as statistical 
inferences from spectroscopy, support some flattening of CE in Be stars (Arias 
et al. 2006; Quirrenbach et al. 1997; Tycner et al. 2005). In the frame of 
theoretical inferences of CE properties based on continuing mass ejections, 
the thinness of circumstellar discs in Be stars still remains an open issue 
(Ara\'ujo \& Freitas Pacheco 1989; Bjorkman \& Cassinelli 1993; Chen et al. 
1992; Owocki et al. 1996; Poe \& Friend 1986; Porter 1997;  Stee \& Ara\'ujo 
1994; Stee et al. 1995, 1998). Moreover, from physical first 
principles (Schmitz 1983) and the kinetic theory of gases applied to steady-state
isothermal self-gravitating gaseous masses with cylindrical differential 
rotation, Rohrmann (1997) predicts bagel-shaped CE that are not reminiscent of
flat discs.\par
2) \textsc{Discrete mass ejections}. In addition to these continuing mass ejections, there are discrete mass ejections. They can be 
divided roughly into two types: 1) small-scale to moderate ejections, detected either as sharp absorption features crossing the line profiles or 
as a sudden appearance of emission shoulders in the wings of some 
absorption lines and/or H$\alpha$ line emission outbursts that imply ejecta 
with masses $\Delta{\rm M}\!\la$ $10^{-11}$M$_{\odot}$ (cf. Floquet et al. 
2000); 2) Large-scale or massive ejections. A clear example is $\omega$~Ori,
where a sudden light increase of nearly $\Delta V\!\approx$ $-0.3$ mag was 
reported by Guinan \& Hayes (1984). The polarimetric data and H$\alpha$ 
spectroscopy of the event were interpreted by Brown \& Wood (1992) as due to a 
discrete mass ejection equivalent to an instantaneous mass-loss rate 
$\dot{M}\!\approx$ $2.5\!\times\!10^{-8}M_{\odot}$~yr$^{-1}$. The sudden 
H$\alpha$ emission outbursts in $\mu$~Cen on timescales ranging from 20 to 
245 days were interpreted by Hanuschik et al. (1993) as being produced by discrete 
mass ejecta of $\Delta M\!\approx$ $10^{-10}M_{\odot}$. In the past decade, 
short- and long-lived photometric outbursts have been discovered (Cook et al. 1995; 
Hubert \& Floquet 1998; Keller et al. 2002; Mennickent et al. 2002) from {\sc
macho}, {\sc ogle}, and {\sc hipparcos} data. They are currently called 
`flickers' or 'sharp' variations when the time scales of phenomena range from
$t\sim$ 50 to 200 days, and they imply light changes of $\Delta V\!\la$ 0.2 mag. 
They are called `bumpers' or `humps' if $t\ga$ 200 days to 3-4 years and 
$\Delta V\!\ga$ 0.3 mag. Mennickent et al. (2002)  speculate that thermal
instabilities may produce these photometric behaviors, while Moujtahid et al. 
(1999), Hubert et al. (2000), and Zorec et al. (2000a) have attempted to explain 
them in terms of sudden or discrete mass-ejections of $\Delta M\!\approx\!
10^{-10}M_{\odot}$. These variations also summarize the irregular photometric
behaviors, including the well-known 
B$\rightleftharpoons$Be$\rightleftharpoons$Be-shell phase transitions, 
observed in the past more or less sporadically (cf. Pavlovski et al. 1997, Percy 
et al. 1997, Moujtahid et al. 1998). Finally, we should not omit mentioning 
the far-UV C\,{\sc iv}, Si\,{\sc iii}, and Si\,{\sc iv} discrete line-absorption components in stars with $V\!\sin i\ga$ 150 km~s$^{-1}$ (Grady et 
al. 1987, 1989), which were interpreted as due to wind instabilities, but  
may actually correspond to discrete mass-ejections.\par 
 According to these observational grounds for mass ejections, two scenarios 
can be sketched to roughly account for the CE variations and/or formation, to 
which the observed long-term Be star emission 
variations should be associated.\par
 This paper is organized as follows. In Sect~2 we introduce the two possible scenarios for the mass ejection. In Sect.~3 we briefly present the SIMECA
code and its main assumptions. Section 4 presents the {\it ring scenario} and Sect.~5  treat the {\it vanishing mass flux scenario}. The comparison of 
different observables (photometric, spectroscopic, and interferometric) is done 
in Sect.~6. Finally, in Sect.~7 we summarize and discuss our main results.\par
\medskip
\section{Two possible scenarios for the mass ejection}

\begin{figure*}[t]
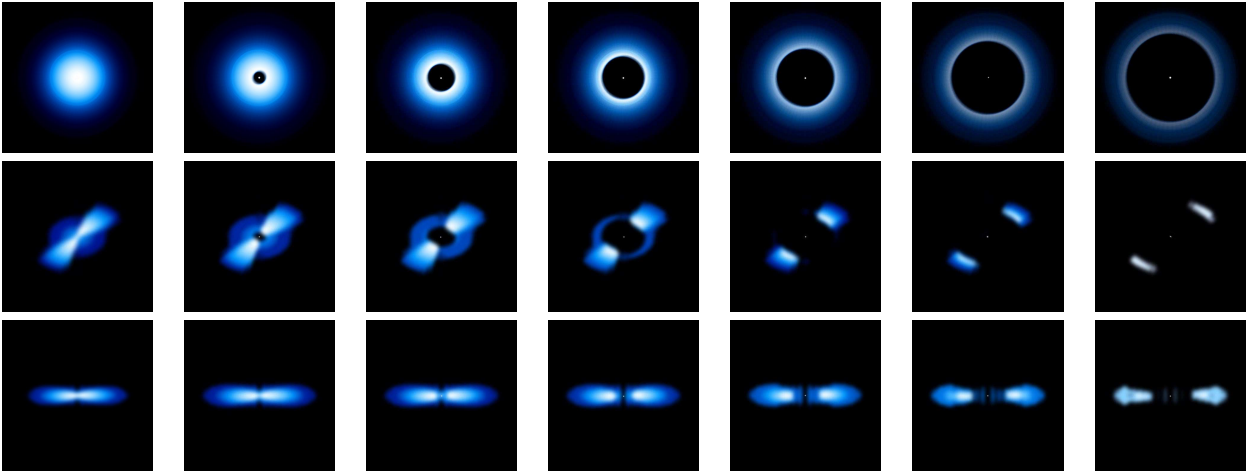

\begin{center}
\begin{tabular}{ccccccc}
\includegraphics[width=2cm,height=2cm]{4690F002.epsf}& 
\includegraphics[width=2cm,height=2cm]{4690F003.epsf}&
\includegraphics[width=2cm,height=2cm]{4690F004.epsf}&
\includegraphics[width=2cm,height=2cm]{4690F005.epsf}&
\includegraphics[width=2cm,height=2cm]{4690F006.epsf}&
\includegraphics[width=2cm,height=2cm]{4690F007.epsf}&
\includegraphics[width=2cm,height=2cm]{4690F008.epsf}\\
\includegraphics[width=2cm,height=2cm]{4690F009.epsf}&
\includegraphics[width=2cm,height=2cm]{4690F010.epsf}&
\includegraphics[width=2cm,height=2cm]{4690F011.epsf}&
\includegraphics[width=2cm,height=2cm]{4690F012.epsf}&
\includegraphics[width=2cm,height=2cm]{4690F013.epsf}&
\includegraphics[width=2cm,height=2cm]{4690F014.epsf}&
\includegraphics[width=2cm,height=2cm]{4690F015.epsf}\\
\includegraphics[width=2cm,height=2cm]{4690F016.epsf} &
\includegraphics[width=2cm,height=2cm]{4690F017.epsf}&
\includegraphics[width=2cm,height=2cm]{4690F018.epsf}&
\includegraphics[width=2cm,height=2cm]{4690F019.epsf}&
\includegraphics[width=2cm,height=2cm]{4690F020.epsf} &
 \includegraphics[width=2cm,height=2cm]{4690F021.epsf}&
\includegraphics[width=2cm,height=2cm]{4690F022.epsf}\\
\end{tabular}
\caption{Intensity maps computed with the SIMECA code showing the formation of a ring ranging from 0 to 60 stellar radii by 10 R$_{*}$ steps. Seen pole-on (upper row), at 45 $\degr$ (center), and equator-on (lower row)}
\label{intensitymaps}
\end{center}
\end{figure*}

\begin{figure*}[t]
\begin{center}
\begin{tabular}{ccc}
\includegraphics[width=5cm,height=5cm]{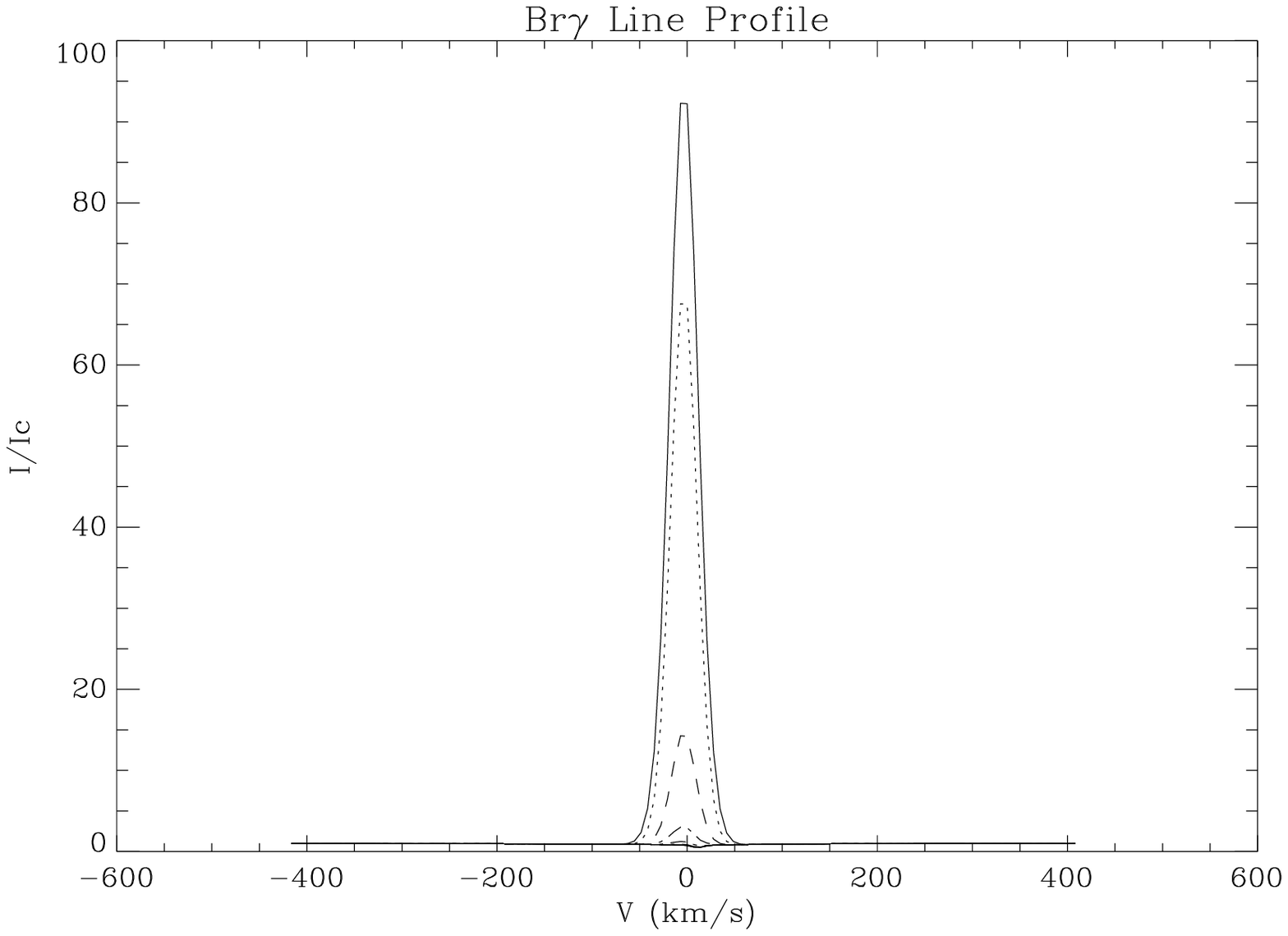}&
\includegraphics[width=5cm,height=5cm]{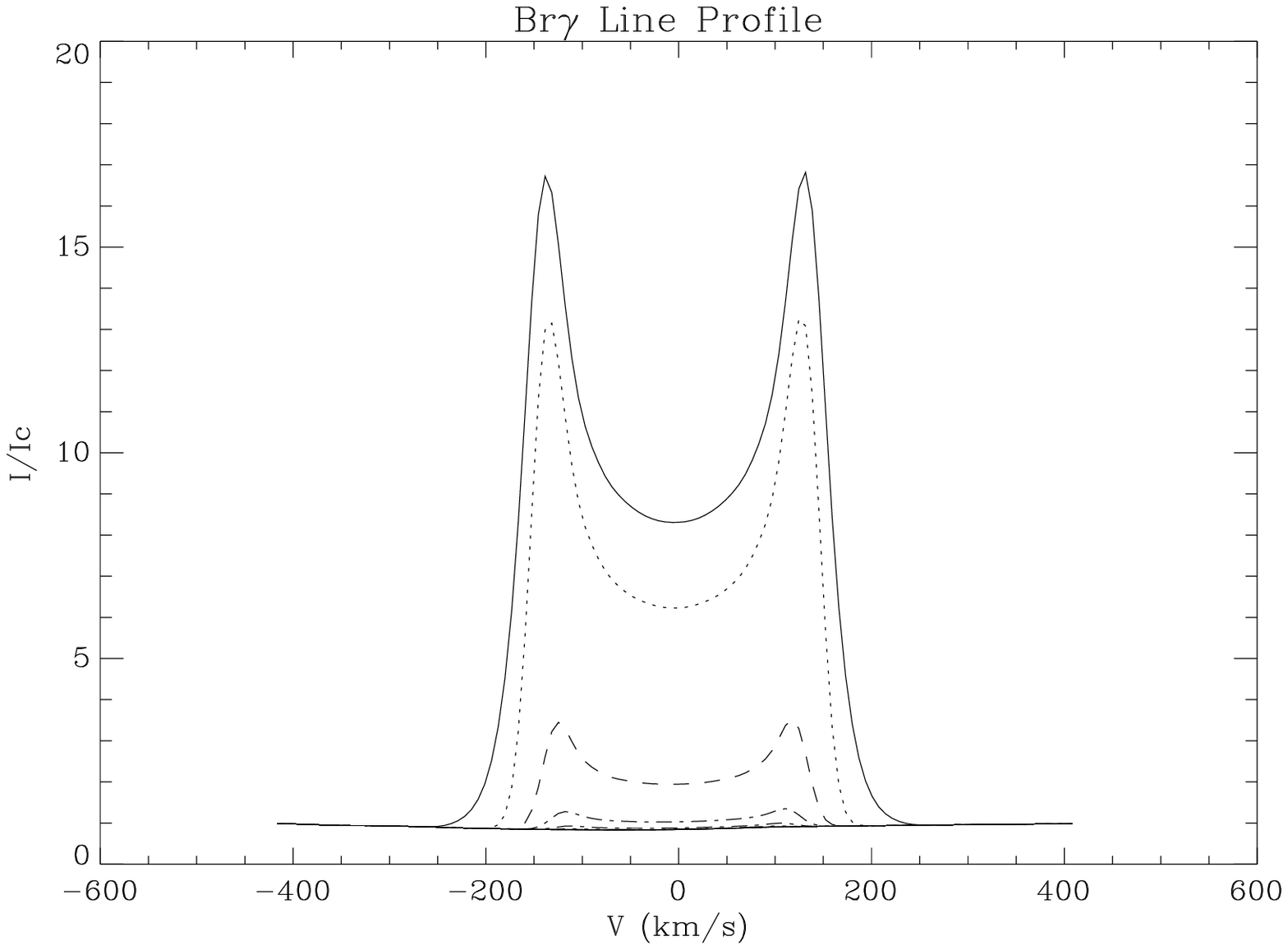}&
\includegraphics[width=5cm,height=5cm]{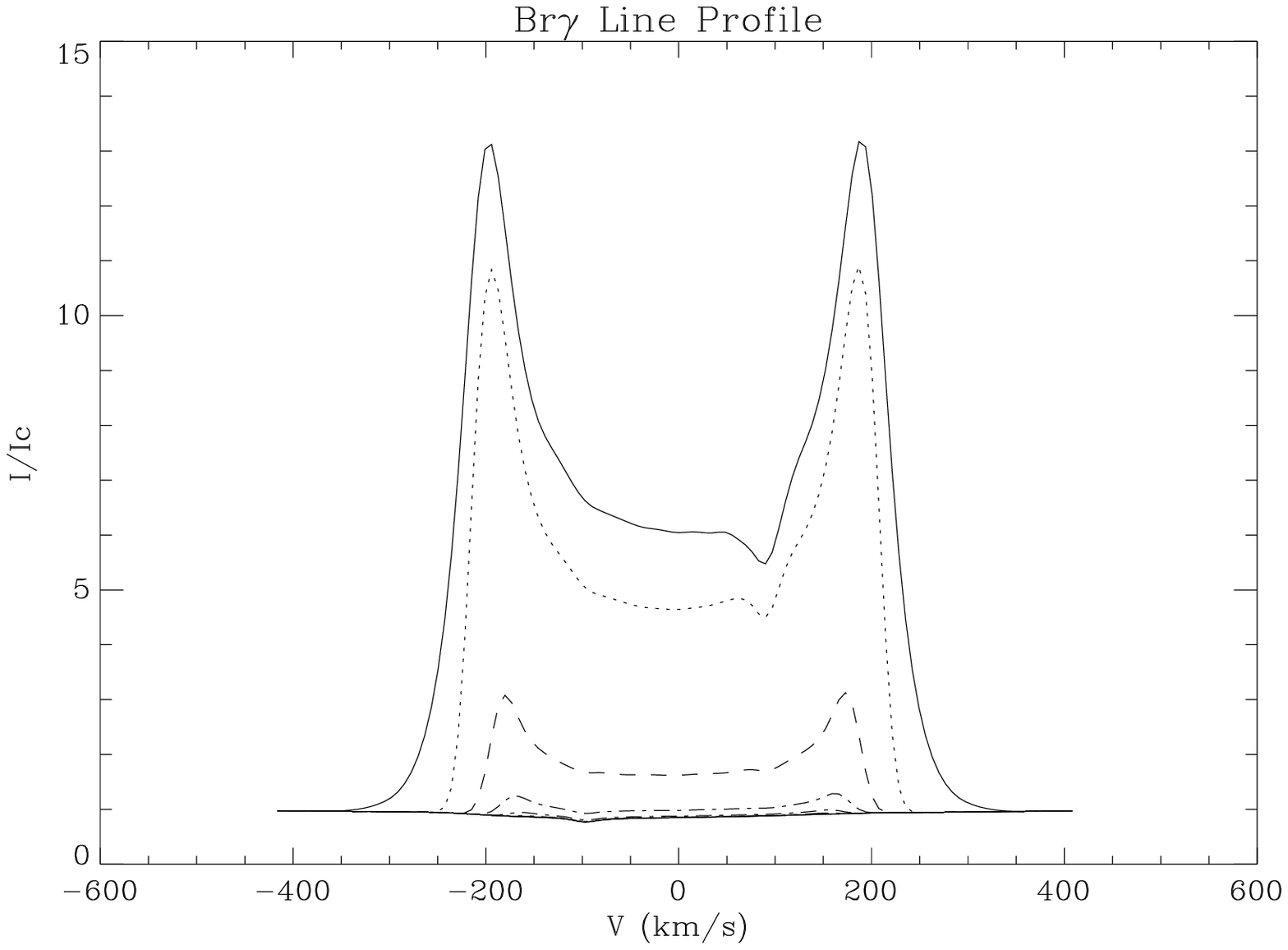}\\
\includegraphics[width=5cm,height=5cm]{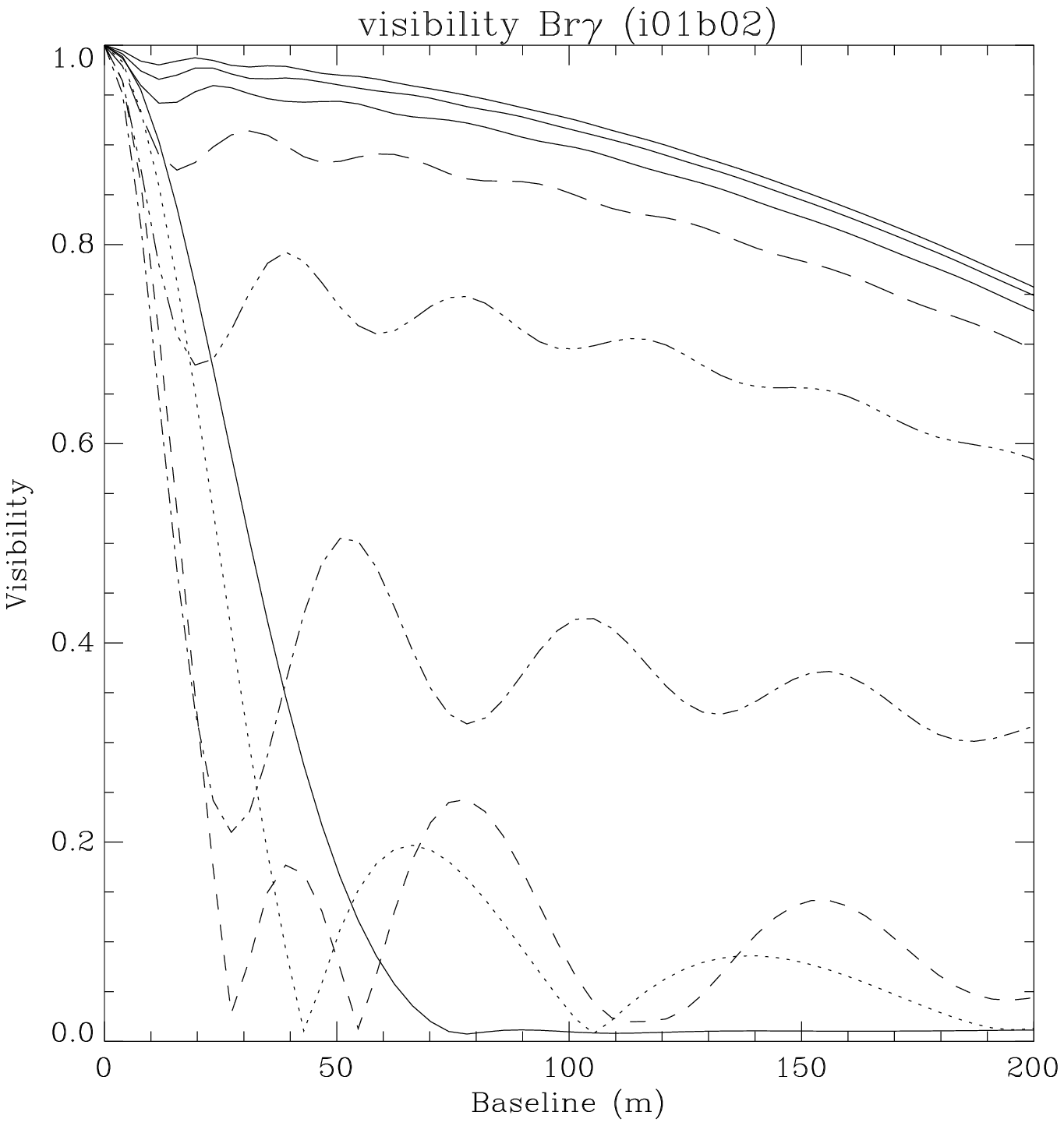}&
\includegraphics[width=5cm,height=5cm]{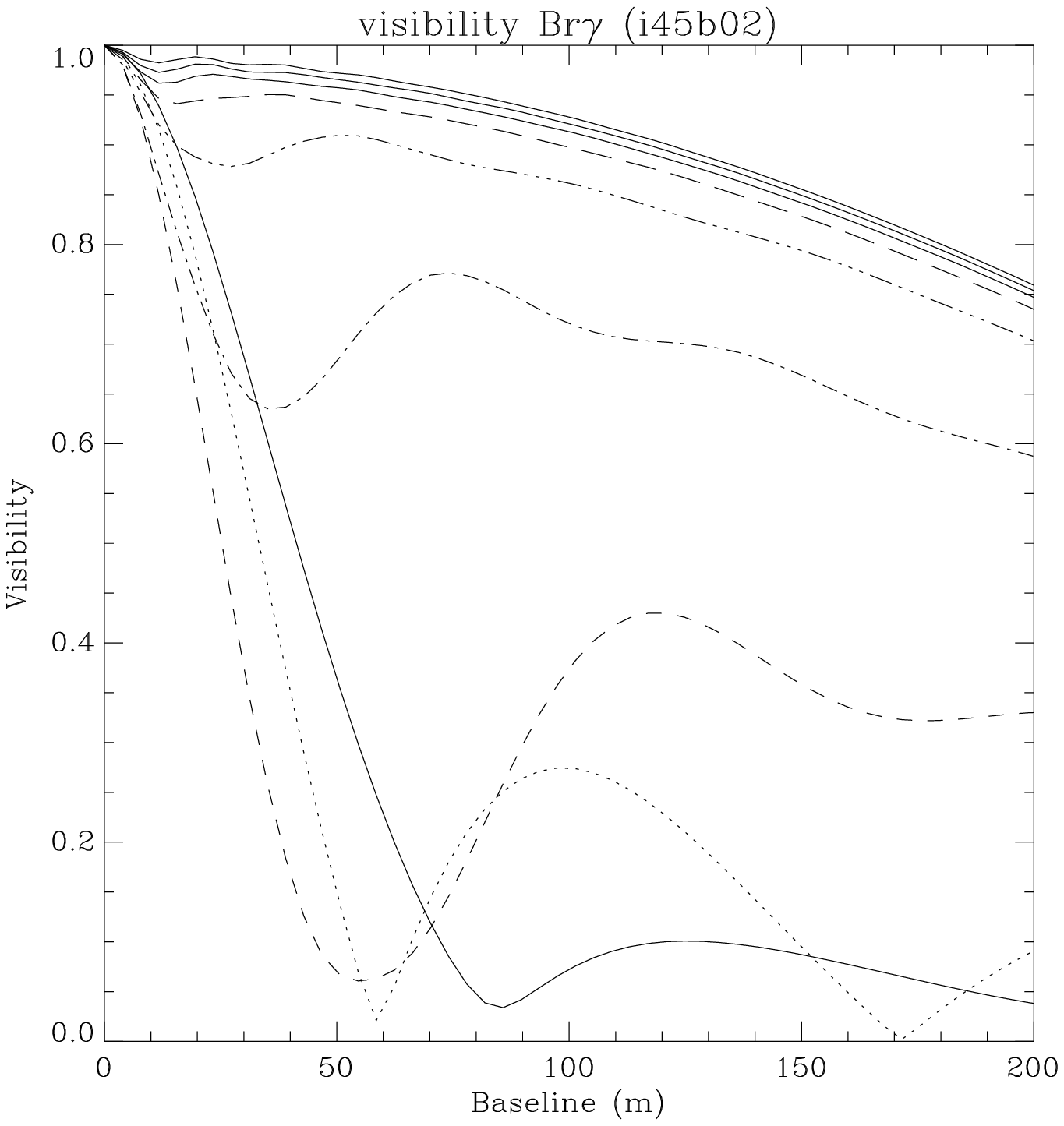}&
\includegraphics[width=5cm,height=5cm]{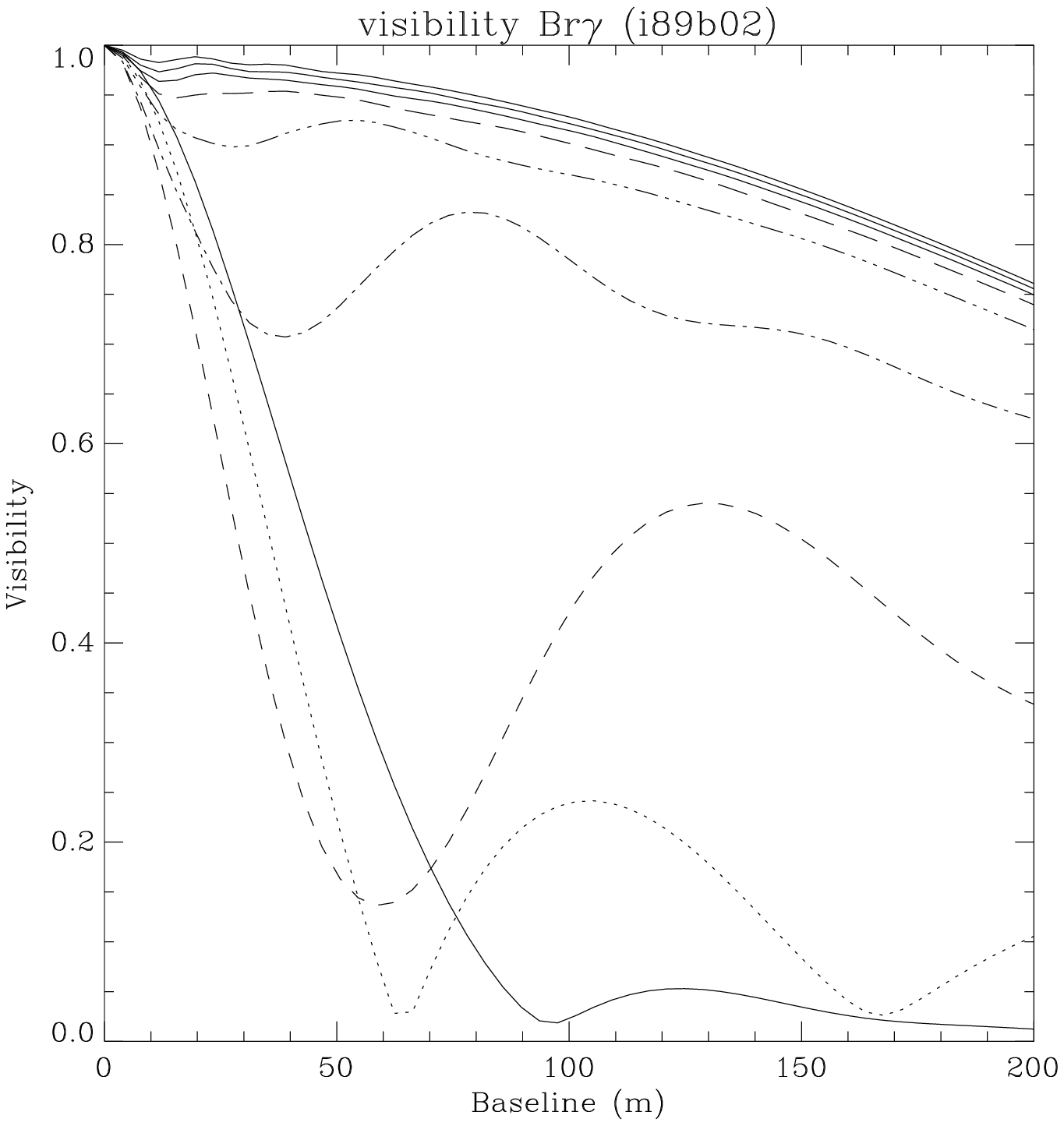}\\
\end{tabular}
\caption{Br$\gamma$ line profiles (upper line) and visibilities as a function of baseline (lower line) for different  ring sizes: r=0 (plain line), r=10 (dotted line), r=20 (dashed line), r=30 (dash-dotted line), r=40 (dash-dot-dot-dot line), r=50 (long-dash line), and for r= 60, 70, and 80 R$_{*}$(plain line) corresponding to the maps given in Fig. \ref{intensitymaps}. These curves are plotted for 3 inclination angles, i.e. Pole-on (left), 45$\degr$ (center),  equator-on (right), and for a baseline position along the equatorial disk.}
\label{lines_visibilities_ring}
\end{center}
\end{figure*}

 \begin{figure*}
  \begin{center}
     \includegraphics[width=7cm,height=7cm]{4690F028.ps}
     \includegraphics[width=7cm,height=7cm]{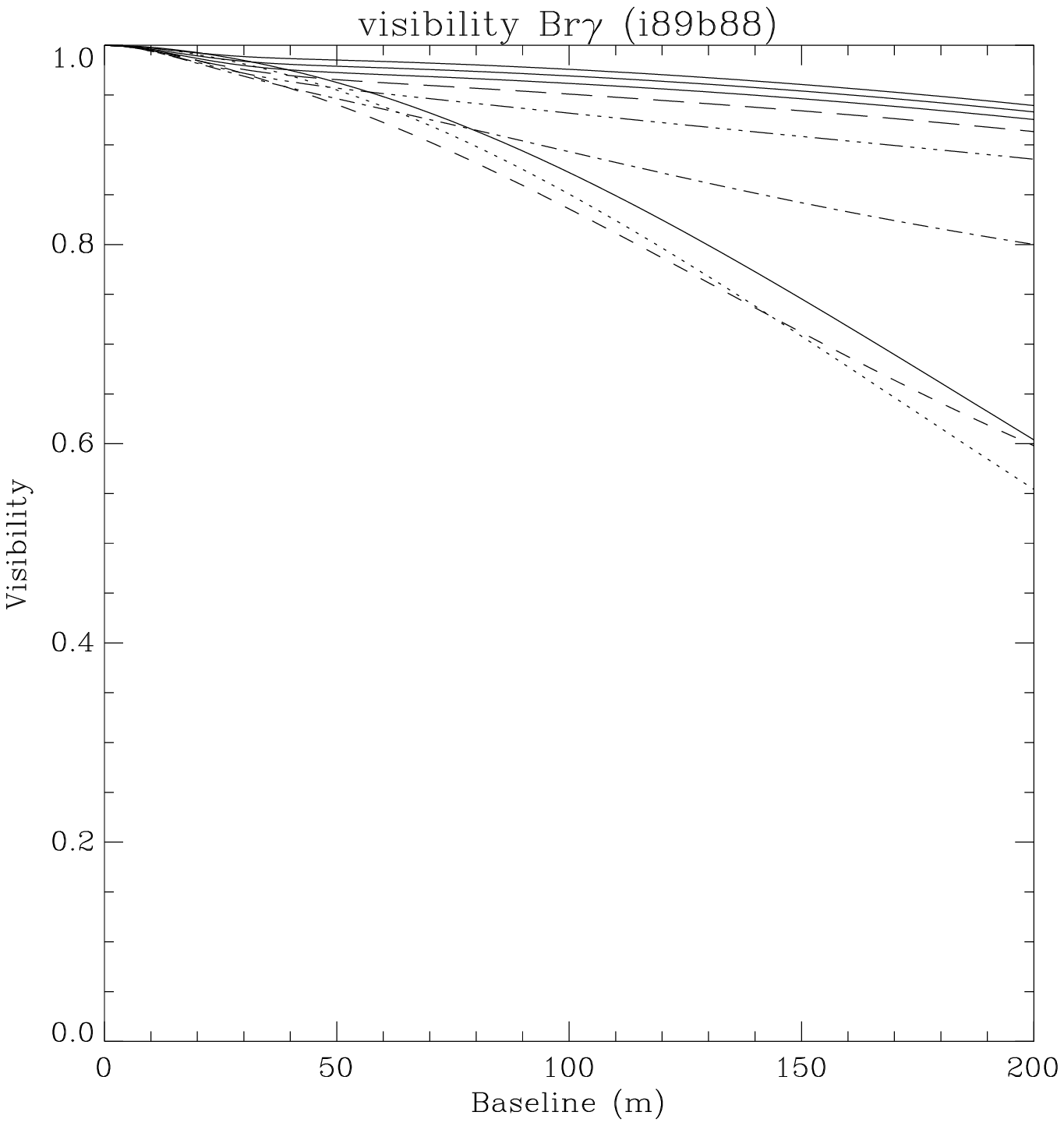}
  \end{center}
  \caption{Visibilities for the ring model as a function of the baseline length for a baseline orientation along (left) and perpendicular (right) to the equatorial disk as a function of the disk dissipation.}
\label{visi_para_perp}
\end{figure*}

 \textsc{Scenario 1: \underline{Expanding rings}}\par
\medskip
 The stellar wind interacts with the {\sc ism} and the matter strewn by 
earlier mass-loss events to create a propagating `shell'-front. If we neglect details for shocks, compression, heating, and cooling phenomena in 
the wind+close interstellar interaction region (Pikel'ner 1986, Kwok et al. 
1978), the balance between the original stellar wind momentum against that of
the wind+swept {\sc ism} and spread material gives the phase with mass-flux 
$\dot{M}_*$ (Zorec 1981) for the location of the `shell'-front at the time $t$ since the star began :
\begin{equation}
R/R_o = [(V_*/v-1)(N_o/N_{\rm ISM}(3V_*t/R_o)]^{1/3},
\label{wsh}
\end{equation}
\noindent where $R_o$ is the radius at which the wind has maximum velocity 
$V_*$, and the particle concentration is $N_o$; $v$ is the expansion of the 
shell'-front; $N_{\rm ISM}\!\approx$ 1 cm$^{-1}$ is the density in the {\sc 
ism}. The lower limit on $R/R_o$ is set by $V_*t=R$. Then $R/R_o\!\approx\!10^4$ for a B2e star, 
taking $R_o\!\approx\!3R_*$ with $R_*\!\approx$ $6R_{\odot}$, $\dot{M}_*\!
\approx$ $10^{-10}M_{\odot}$yr$^{-1}$ (Snow 1982), $V_*\!\approx$ 500 
km~s$^{-1}$, $v/V_*\!\approx 0.1$. Such
a 'shell', occurring at $3\!\times\!10^4R_*$ after 8 years of steady mass-flow,
can hardly satisfy the characteristics required by CEs in Be stars. However, 
the mass-loss phenomenon is variable. Zorec (1981) suggested then that 
long-lasting low mass-loss episodes can build up moving bubbles around the 
star, which may act as dams to pile up matter ejected during later 
short high mass-loss events and thus increase the density around the 
star up to values typical of CE. To this view Rivinius et al. (2001) have 
joined a roughly similar picture, which suggested the appearance of a growing 
cavity between the star and an already existing CE, which is
replenished later on with the mass of new ejecta. According to these suggestions, the 
variable emission intensity in Be stars might then correspond to the 
appearance of a density void-like region around the star and an expanding 
ring-like structure.\par 
 This intuitive phenomenological picture can be justified on physical 
grounds taking both the large-scale discrete ejections and winds into account. 
As seen above, discrete ejections can be effective providers of mass to form CE 
in Be stars. On the other hand, stellar winds can contribute to the 
global dynamics of the mass stored around the star not only with 
non negligible amounts of mass, but also with momentum and energy. The CE could then be 
thought of as an expanding wind-blown bubble formed by the process of interaction between the stellar wind and massive discrete ejecta: discrete ejecta produce a 
clumpy circumstellar environment, which is then ablated by winds producing a 
mass-loaded flux. The structure of expanding wind-blown bubbles produced by 
the interaction of high-speed stellar winds with clumpy environments were 
studied by Hartquist et al. (1986), Dyson \& Hartquist (1992), and Arthur et 
al. (1994). According to these authors, the most simplified structure of the 
circumstellar environment in the stationary snowplow phase encompasses three 
dynamically distinct regions: 1) a wind expansion region; 2) a decelerated, 
sub-sonic wind momentum-dominated {\it core}; 3) a pressure-dominated 
super-sonic expanding {\it halo}. In the decelerated `core' the density rises 
with the distance as $R^4$ from the radius $R_m$ of minimum environmental 
density up to a contact radius $R_c\!\approx\!(2V_*/v_s)R_m$ ($v_s$ is the 
sound speed) that separates the sonic and super-sonic sides. In the `halo', the 
density decreases as ${\rm e^{-M^2/2}}$ where $M\!=\!v/v_s$ is the local Mach
number in the flow. According to Arthur et al. (1994), $R_m\!=$ 
$(3\dot{M}_*/8\pi\dot{q})^{1/3}$ where $\dot{q}$ is the mass injection rate 
from the cloud ablation process. To obtain orders of magnitude on the bubble 
expansion and the amount of mass gathered in it, we integrated the 
one-dimensional rotationless steady-state equations of motion given by Dyson 
\& Hartquist (1992). Assuming that the ablation of discrete ejecta starts
at $R_m\!\approx\!R_*$ and the expansion of the bubble is dominated by the 
ablated mass, the time-dependent velocity of the bubble is:
\begin{eqnarray}
\dot{R_b} & \simeq & 0.5(V_*R_*)^{1/4}t^{1/4}.
\label{vbub}
\end{eqnarray}
\noindent Within the same approximation, the gathered mass in the bubble from
$R_m$ up to a given external distance $R_e\!\gg\!R_c$ scales~as:
\begin{eqnarray}
M_{\rm CE} & \simeq & (3R_c\dot{M}_*/V_*)[\!1\!+\!(V_*/v_s)]^2\!
(R_e/R_c)^{1.3},
\label{mce}
\end{eqnarray}
\noindent where the exponent 1.3 is due to a power-law representation of the
density distribution in the halo. With the same parameters as are used to estimate
(\ref{wsh}) and ${\rm R_e\approx10R_*}$, $T_{\rm e}\approx$ $10^4$ K, we
find that at $t\sim$ 1 day the bubble expands at $\dot{R_b}\approx$ 180 
km~s$^{-1}$ and at $\dot{R_b}\approx$ 14 km~s$^{-1}$ for $t\sim$ 1 month. 
In the meantime, the radius of the bubble increases from $R_b\la$ $4R_*$ to $\ga$ 
$8R_*$ and the mass stored in it amounts to some $M_{\rm CE}\approx2\times
10^{-9}M_{\odot}$. In current modelings, the total mass in a CE with base 
particle density $N \sim 10^{12}$ cm$^{-3}$ ranges from $10^{-10}M_{\odot}$ to 
$10^{-8}M_{\odot}$, depending on whether the CEs are  
considered flat disc-shaped or spherical, with uniform density distribution or 
varying with radius as $\sim R^{-2}$. It is then seen that the wind-discrete ejecta interaction 
mechanism can store the amount of mass expected in CE of Be stars on short time scales in a ring-like volume around the star. As seen, the expanding 
ring-like structure leaves a cavity behind it and can act as a dam to 
reconstruct new mass-enhanced circumstellar environments with matter ejected 
by later outbursts, as suggested by Zorec (1981) and Rivinius et al. 
(2001).\par 
\medskip
\textsc{Scenario 2: \underline{Variable mass-loss rates}}\par
\medskip
In the preceding discussion we did not consider the problem of constructing 
CE in (or near) Keplerian rotation. Be stars are fast, though under-critical, 
rotators (Fr\'emat et al. 2005). Comparing the angular momentum per unit mass 
in a disc at Keplerian rotation $J=(GM_*R)^{1/2}$ and at a distance $R$ from the star 
to that at the stellar surface, assumed at critical rotation (Keplerian too), $J_*=(GM_*R_*)^{1/2}$,
we see that matter ejected from the star needs at least a factor $J/J_*=
(R/R_*)^{1/2}$ of angular momentum-excess to become Keplerian at $R$. Thus
to recover this rotation in the circumstellar disc, Lee et al. (1991), Porter 
(1999), and Okazaki (2001) envision outward-drifting matter in the 
circumstellar environment through angular momentum exchange by viscous stress. 
Unless hydrodynamic or magneto-hydrodynamic instabilities in the disc induce 
variations in its structure, variable emission intensities can be thought of as triggered only by variable mass decretion rates in the present scenario. 
This viscous decretion disc model seems to describe the time 
scales of the disc formation correctly, as well as its further dissipation in the Be-shell star 
$o$~And (Clark et al. 2003). The time scales needed to build up the CE in 
$\alpha$~Eri, as scaled from the observed long-term H$\alpha$ line emission
increases, also agree with those predicted by this scenario. However, the 
dissipation of its CE requires somewhat longer time scales than foreseen by 
this theory (Vinicius et al. 2006).\\

 According to {\it scenarios 1} and {\it 2}, the increase in 
the emission in Be stars should correspond either to the replenishment of a 
circumstellar cavity by mass input from the star (wind and/or discrete ejecta)
or to a continuous decretion rate that is increased at any moment by some 
mechanism. On the other hand, the fading of emission characteristics would 
correspond either to a cavity formation according to the ring scenario or to
a disc dissipation carried by a stopped or reduced mass-decretion rate.\par  
\medskip
\textsc{Aim and progression of the present work}. With the operational {\sc 
vlti} array, with focal instruments {\sc midi}, and in particular with {\sc amber}, 
which combines spatial and spectral resolution of $10^4$, it is now possible 
to study the Br$\gamma$ formation region, as well as the shape and extension of 
the circumstellar disk with a spatial resolution of few {\it mas}; i.e. the
angular diameter of 1 {\it mas} is for a typical Be star with radius $R_*\!=
\!10R_{\odot}$ situated at 100 pc, while the Br$\gamma$ formation region 
extends to some stellar radii. The main purpose of this paper is then to give
observational grounds for probing the above scenarios of circumstellar disc 
evolution by searching whether, through interferometric measurements in the 
Br$\gamma$ line, it is possible to distinguish which of both scenarios would 
better suited to explaining the long-term variation of the emission characteristics 
in Be stars. We would also like to study the consequences on other observables, 
such as line profiles and spectral energy distributions ({\sc sed}) that may 
carry these scenarios.\par

 In this paper we explore the waning aspect of disks and emissions. Thus, we 
present simulations of the evolution of a cavity from the inner edge to the 
outer regions as would happen in Scenario 1 and, on the other hand, 
calculations of the effects produced by a Be star disk submitted to a decreasing 
mass-loss rate as suited to Scenario 2. To these purposes, we used the {\sc 
simeca} code described in previous papers [see Stee \& Ara\`{u}jo (1994) and 
Stee \& Bittar (2001)]. In particular, we studied the evolution and variation of 
the Br$\gamma$ line profile as a function of time, the changes in the spectral 
energy distribution ({\sc sed}) and those of intensity maps with the 
corresponding visibility curves. As we shall see, the visibility curves are 
very sensitive to the disk geometry, so that they can probe the dissipation of the circumstellar matter with high accuracy. We can then use them to make a distinction 
between {\it scenario 1} and {\it 2}.\par

\begin{figure}
  \begin{center}
      \includegraphics[height=8.cm]{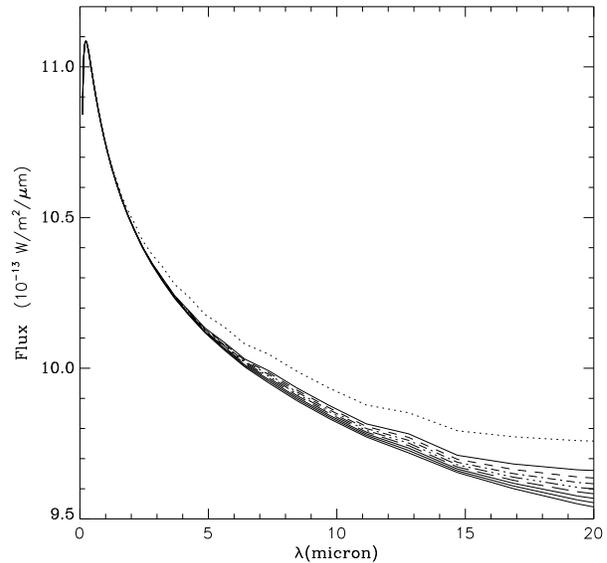}
  \end{center}
  \caption{Spectral energy distribution (SED) as a function of the ring size, for the lower plain line: central star only, r=0 (dotted line), r=10 (dashed line), r=20 (dash-dotted line), r=30 (dash-dot-dot-dot line), r=40 (long-dash line), and for r= 50, 60 70 and 80 (plain line). Since the disk is nearly optically thin the SED is nearly independent of the inclination angle between the star rotation axis and the observer line of sight.}
\label{sed_radius}
\end{figure}

\begin{figure*}[t]
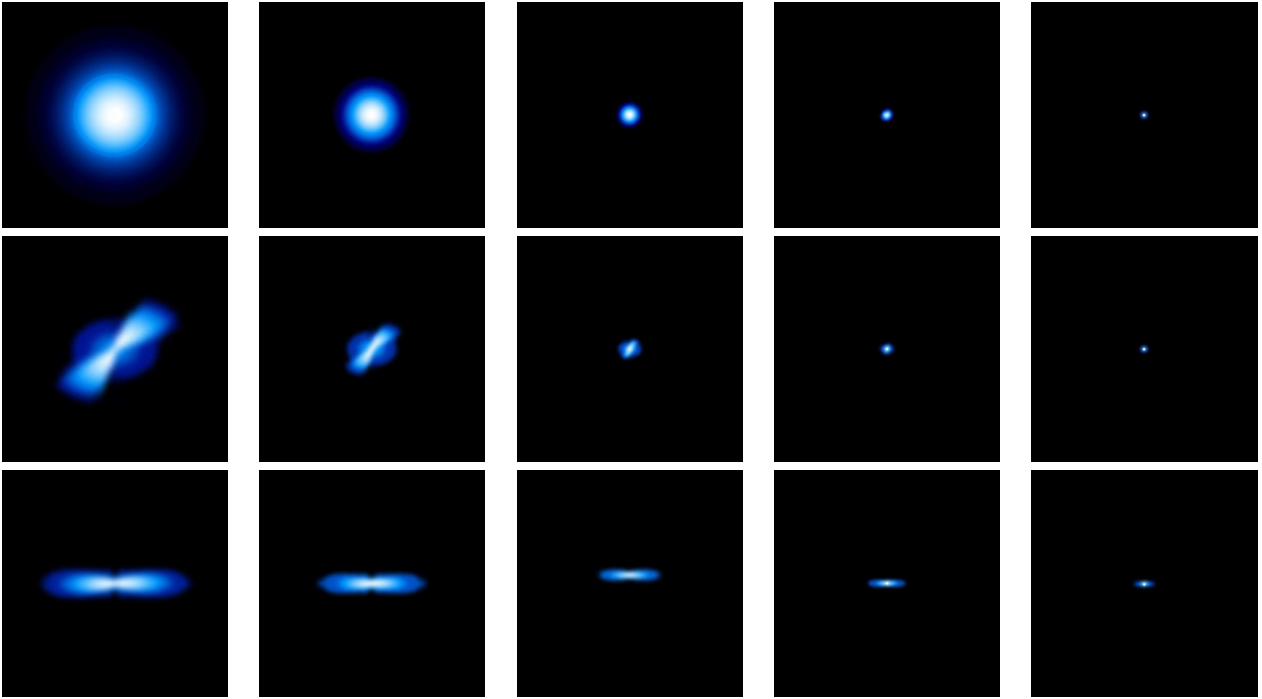

\begin{center}
\begin{tabular}{ccccc}
\includegraphics[width=3cm,height=3cm]{4690F002.epsf} & 
\includegraphics[width=3cm,height=3cm]{4690F031.epsf}&
\includegraphics[width=3cm,height=3cm]{4690F032.epsf}&
\includegraphics[width=3cm,height=3cm]{4690F033.epsf}&
\includegraphics[width=3cm,height=3cm]{4690F034.epsf}\\
\includegraphics[width=3cm,height=3cm]{4690F009.epsf}&
\includegraphics[width=3cm,height=3cm]{4690F035.epsf}&
\includegraphics[width=3cm,height=3cm]{4690F036.epsf}&
\includegraphics[width=3cm,height=3cm]{4690F037.epsf}&
\includegraphics[width=3cm,height=3cm]{4690F038.epsf}\\
\includegraphics[width=3cm,height=3cm]{4690F016.epsf}&
\includegraphics[width=3cm,height=3cm]{4690F039.epsf}&
\includegraphics[width=3cm,height=3cm]{4690F040.epsf}&
\includegraphics[width=3cm,height=3cm]{4690F041.epsf}&
\includegraphics[width=3cm,height=3cm]{4690F042.epsf}\\
\end{tabular}
\caption{Intensity maps computed with the SIMECA code showing the vanishing of the circumstellar disk by decreasing the mass flux and thus the global mass loss rates:  4.7 10$^{-10}$, 2.3 10$^{-10}$, 9.4 10$^{-11}$, 4.7 10$^{-11}$, 2.3 10$^{-11}$M$_{\odot}$ year$^{-1}$ from the left to the right, as seen pole-on (upper row), at 45 $\degr$ (center), and equator-on (lower row) }
\label{intensitymaps_flux}
\end{center}
\end{figure*}

\begin{figure*}[t]
\begin{center}
\begin{tabular}{ccc}
\includegraphics[width=5cm,height=5cm]{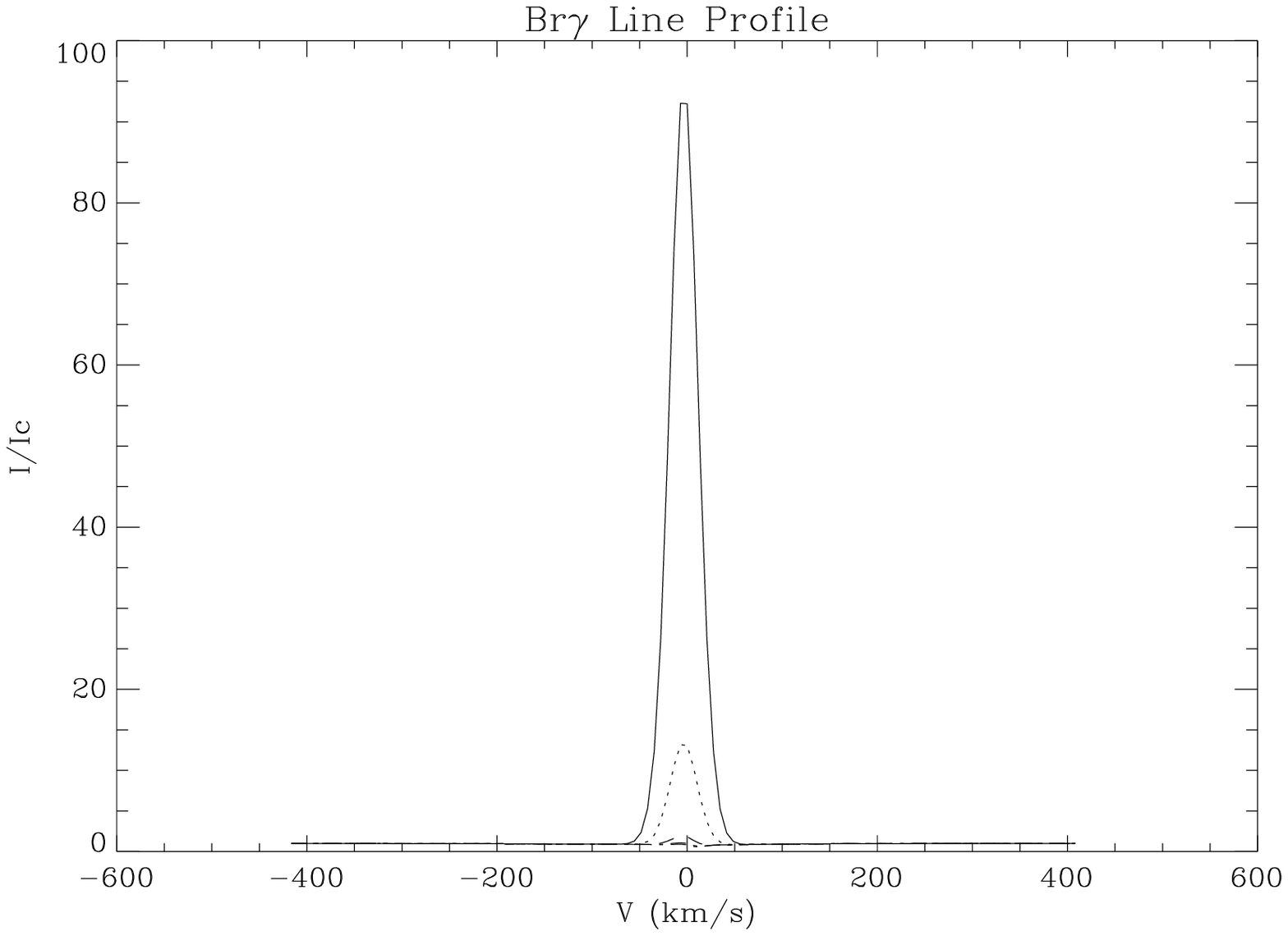}&
\includegraphics[width=5cm,height=5cm]{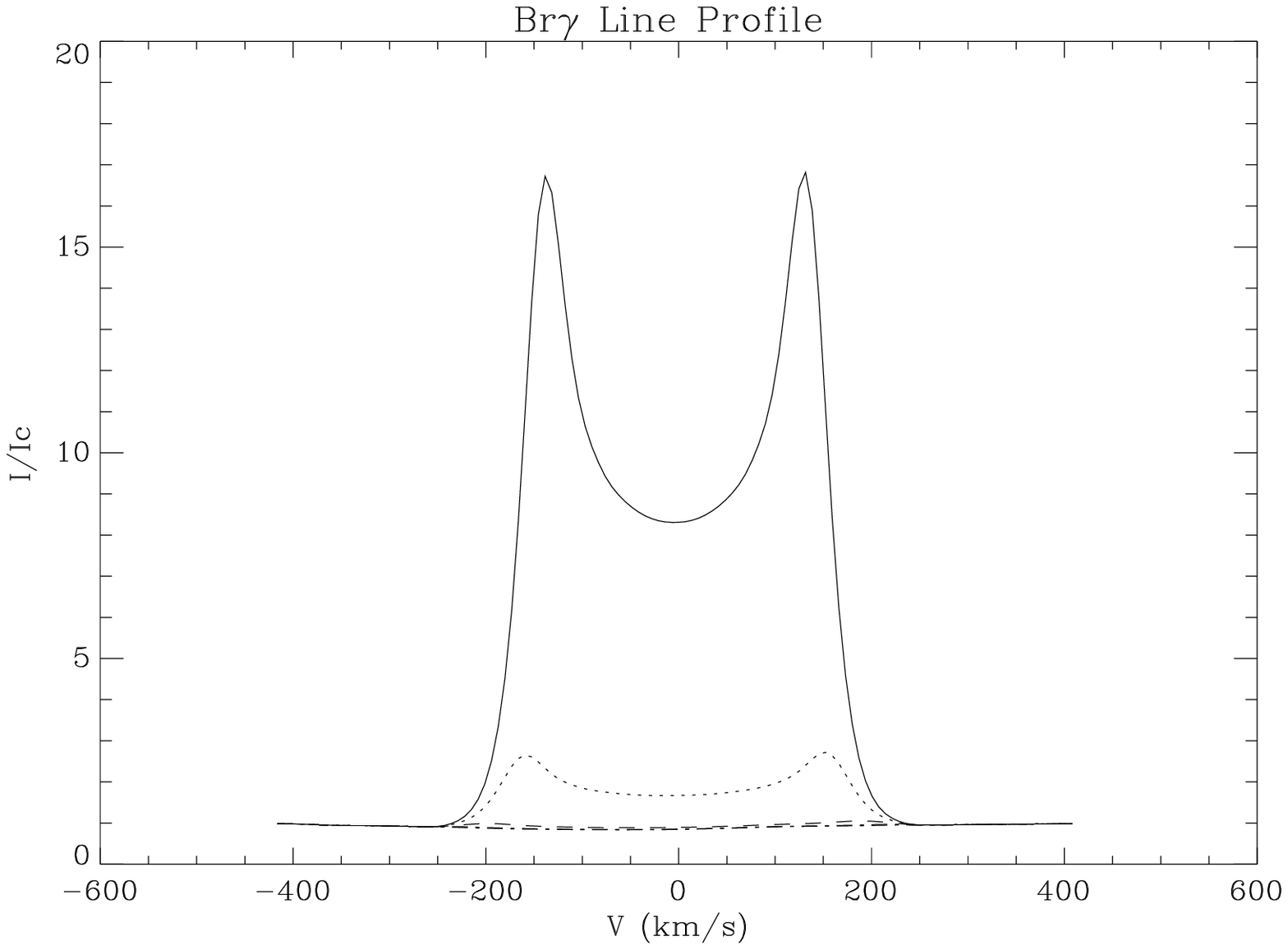}&
\includegraphics[width=5cm,height=5cm]{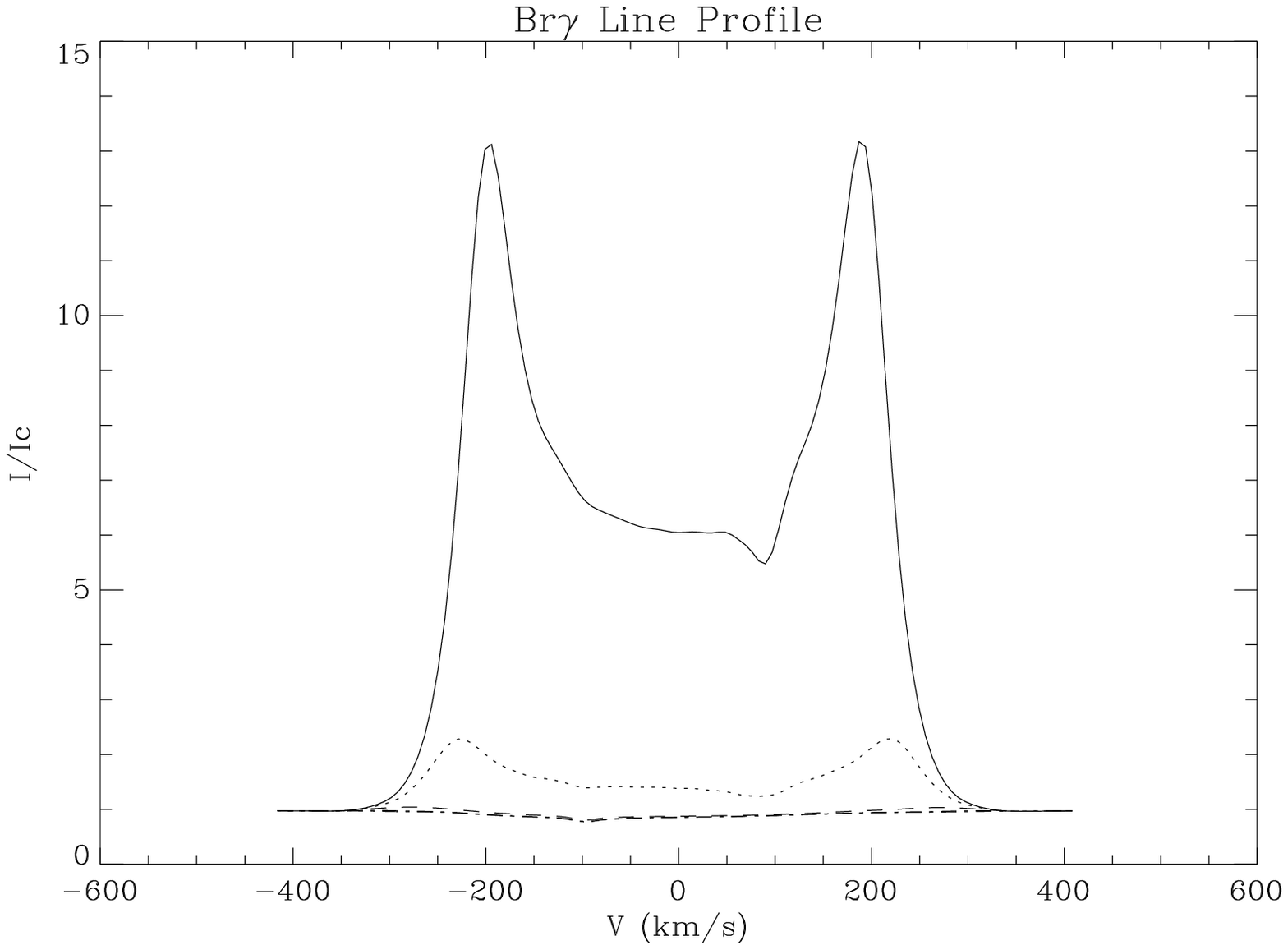}\\
\includegraphics[width=5cm,height=5cm]{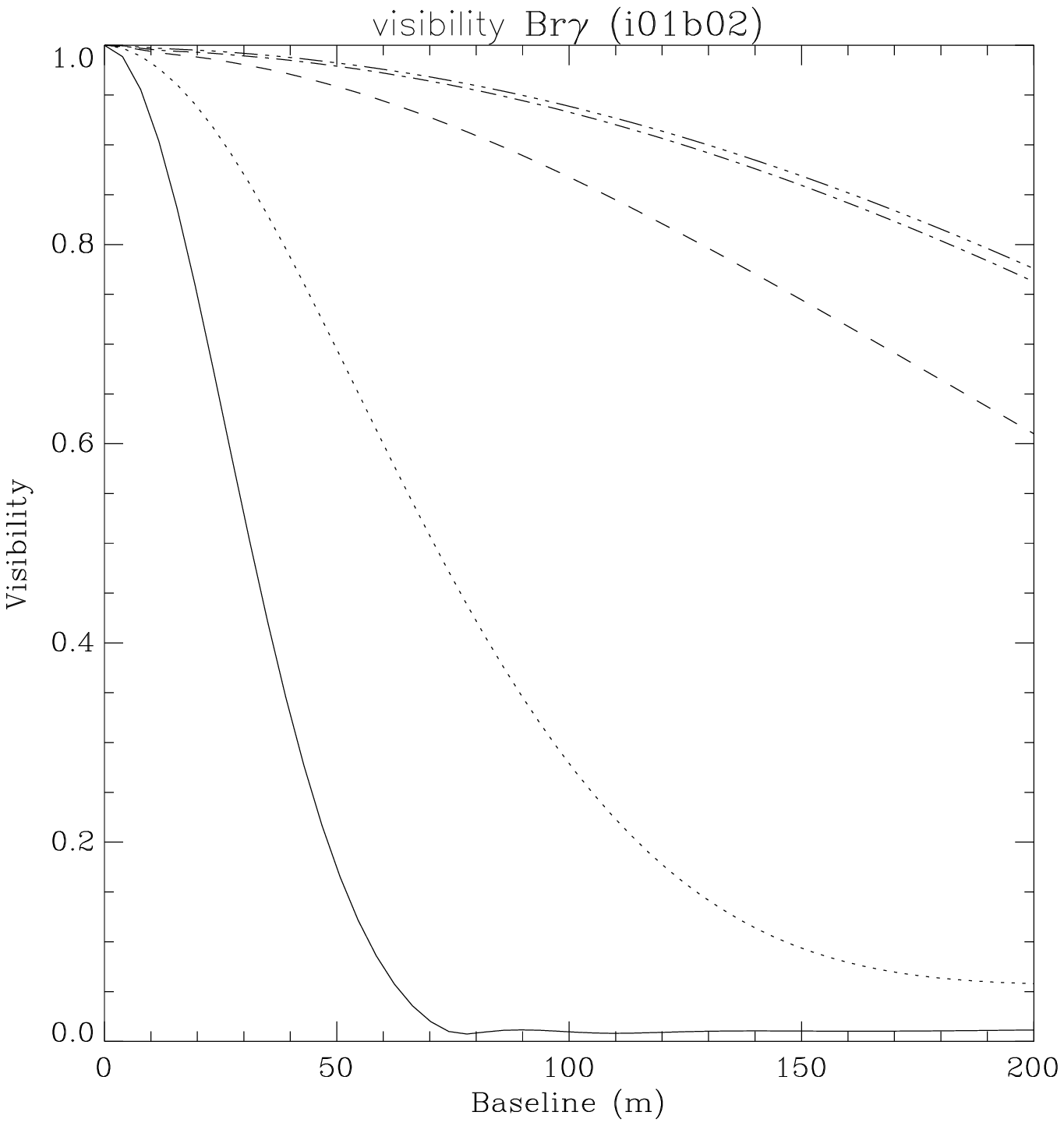}&
\includegraphics[width=5cm,height=5cm]{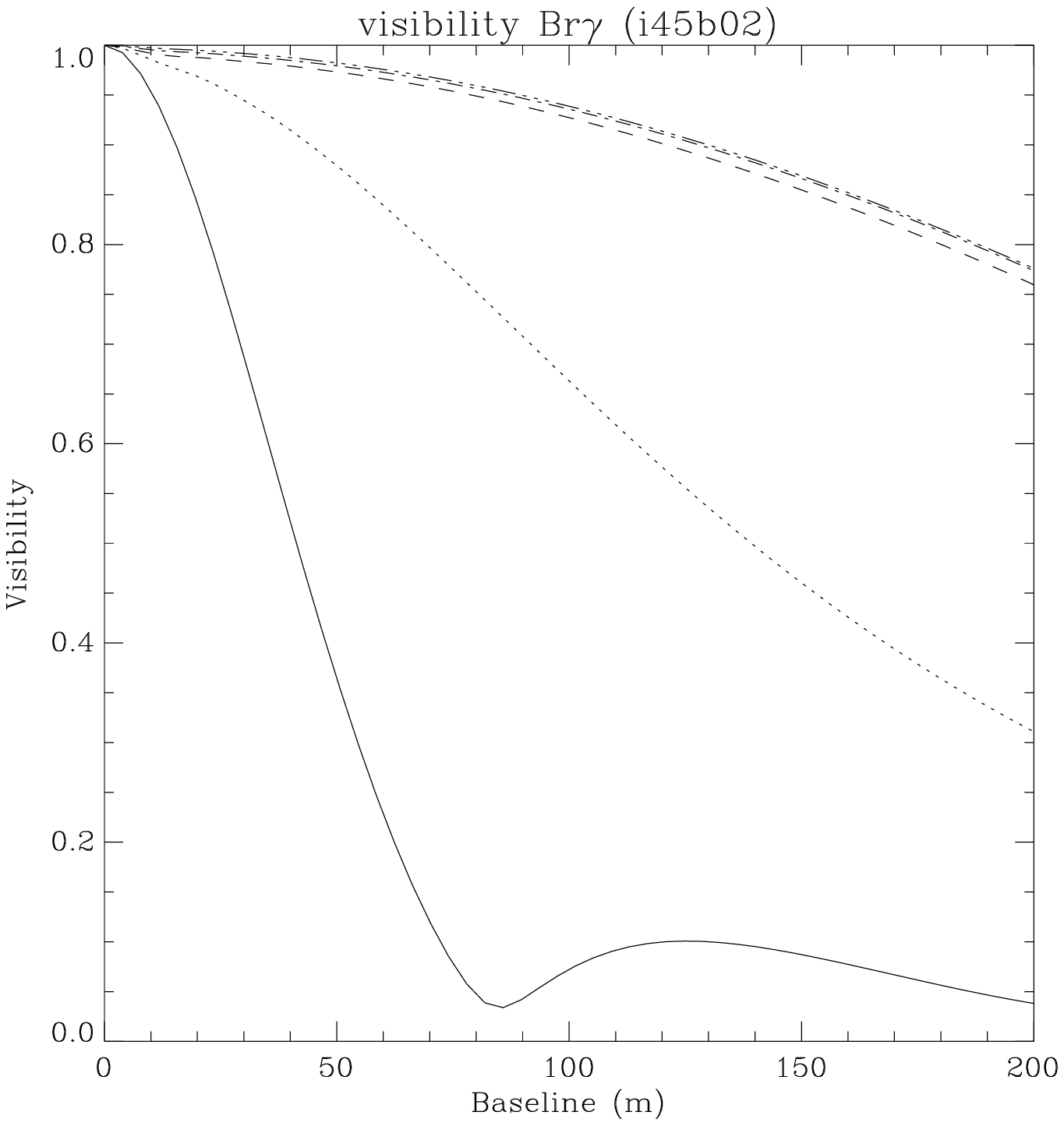}&
\includegraphics[width=5cm,height=5cm]{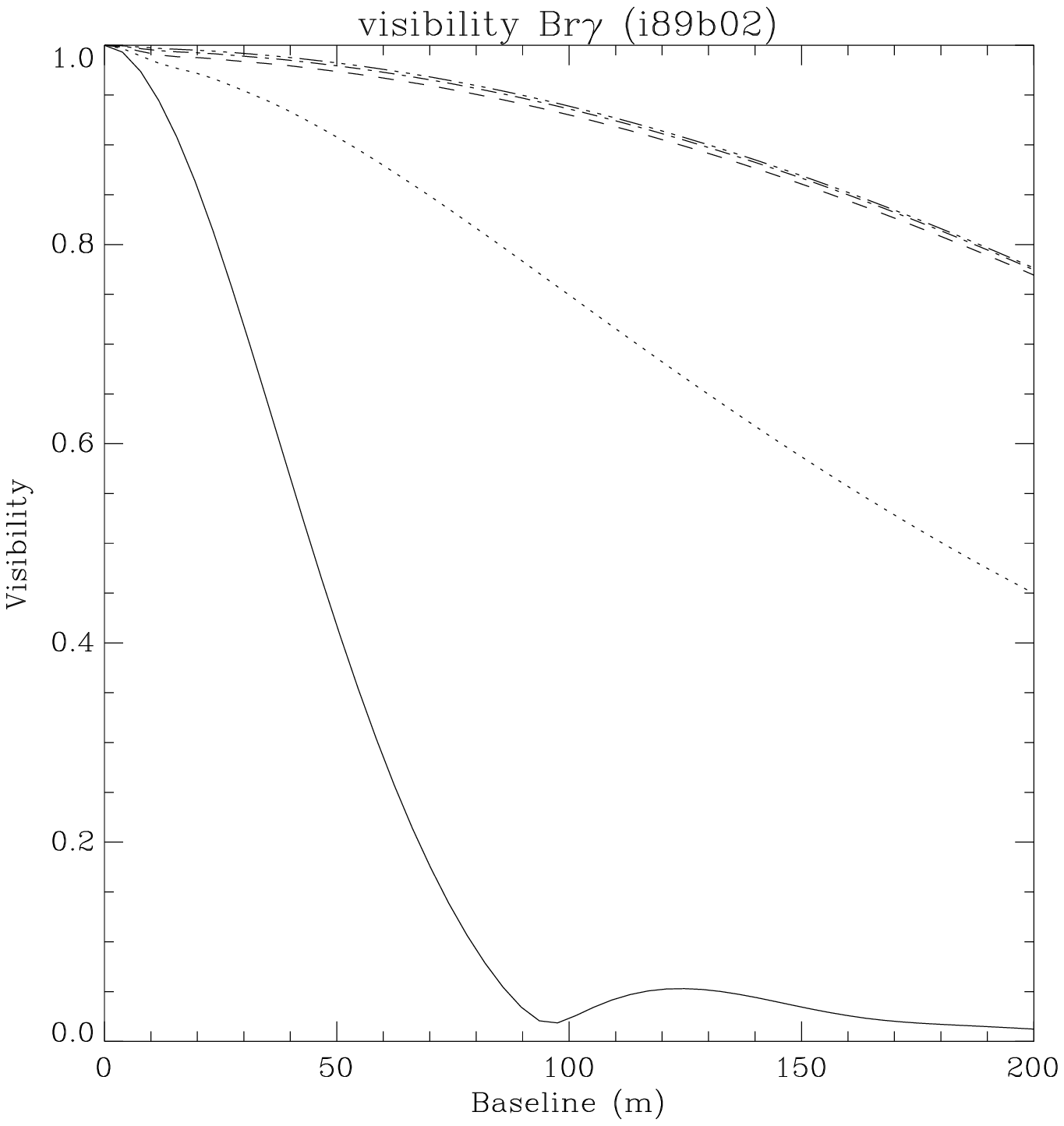}\\
\end{tabular}
\caption{Br$\gamma$ line profiles (upper row) and visibilities as a function of baseline (lower row) for different mass loss rates, 4.7 10$^{-10}$ (plain-line), 2.3 10$^{-10}$ (dotted-line), 9.4 10$^{-11}$ (dashed-line), and 4.7 10$^{-11}$ M$ _{\sun } $year$^{-1}$ corresponding to the maps given in figure \ref{intensitymaps_flux}. These curves are plotted for 3 inclination angles, i.e. Pole-on (left), 45$\degr$ (center),  equator-on (right) and for a baseline position along the equatorial disk.}
\label{lines_visibilities_flux}
\end{center}
\end{figure*}

\section{SIMECA: a code dedicated to active hot stars}

\begin{figure}[t]
  \begin{center}
      \includegraphics[height=8.cm]{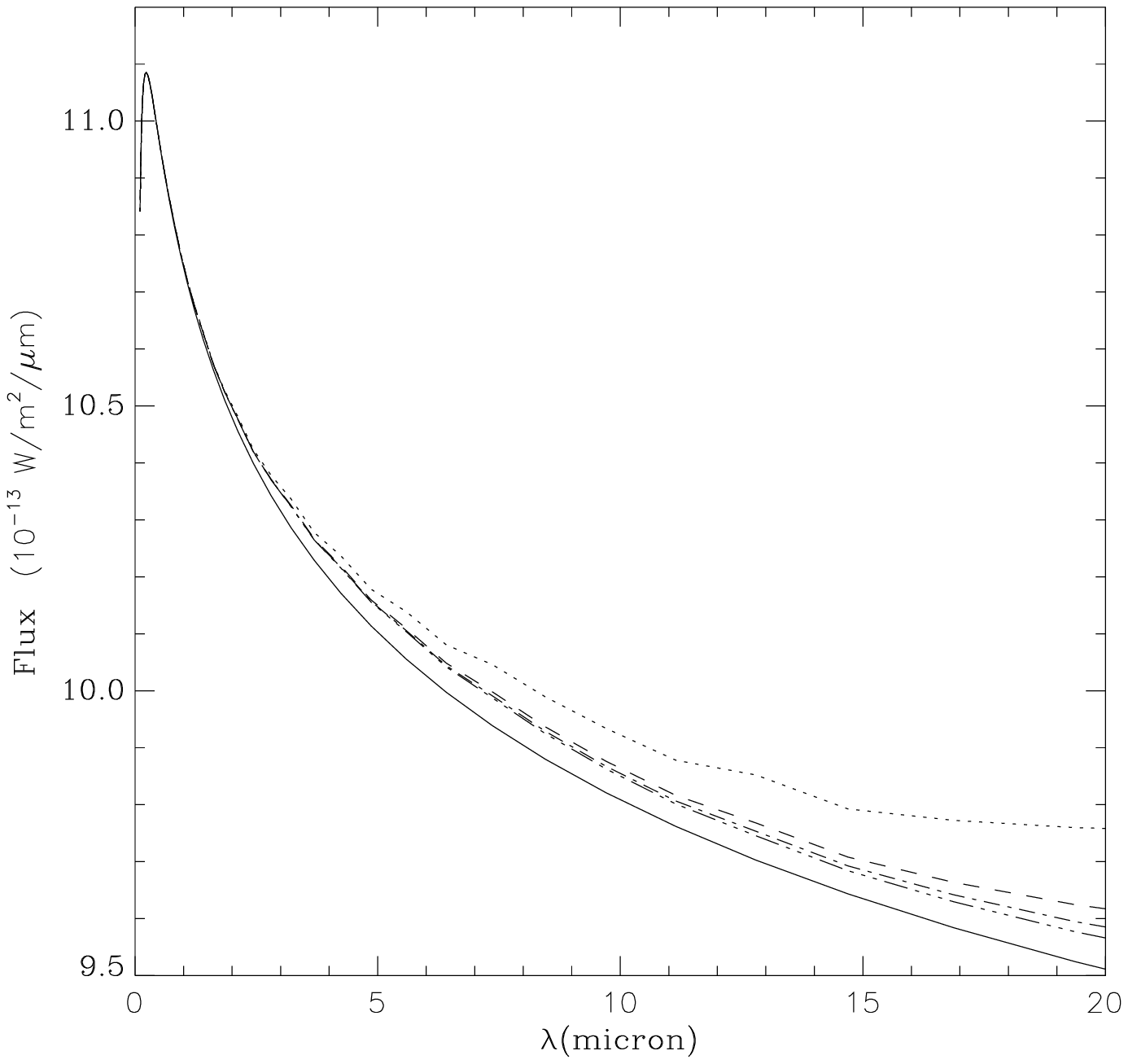}
  \end{center}
  \caption{Spectral energy distribution (SED) as a function of the mass loss rate: only the central star (lower plain line), 4.7 10$^{-10}$ (dotted-line), 2.3 10$^{-10}$ (dashed-line), 9.4 10$^{-11}$ (dash-dot-line) and 4.7 10$^{-11}$ (dash-dot-dot line) M$ _{\sun }$ year$^{-1}$. Since the disk is nearly optically thin, the SED is nearly independent of the inclination angle between the star rotation axis and the observer's line of sight.}
\label{sed_flux}
\end{figure} 

The SIMECA code computes classical observables, i.e. spectroscopic
and photometric ones, but also intensity maps in Balmer lines and in the continuum, in
order to obtain theoretical visibility curves that can be directly compared to high
angular resolution data. The main hypothesis of this code is that the envelope is 
axi-symmetric with respect to the rotational axis. No meridian circulation is allowed. We
assumed that the physics of the polar regions is represented well by a CAK type
stellar wind model (Castor et al. \citealp{Castor}) and the solutions for all stellar
latitudes were obtained by introducing a parametrized model that is constrained
by spectrally resolved interferometric data. The inner equatorial region is
dominated by Keplerian rotation. 

The ionization-excitation equations were solved for an envelope modeled 
in a 411{*}92{*}73 cube using spherical coordinates, for $r$, $\theta$ and $\phi$. 
Since the final population of atomic levels were strongly NLTE distributed, we started
with the LTE populations for each level, then computed the escape probability
of each transition, which allowed us to obtain updated populations, and we iterated
until convergence. The convergence was quite fast (about ten iterations) and
stable within an effective temperature of the central star in the range 10000
$<$ \( T_{eff} \) $<$ 40000. The basic equations of the SIMECA code are given in
detail in Stee \& Ara\`{u}jo \cite{stee0}. \\

To account for the photospheric absorption line, we assumed the underlying star
to be a normal B star with a given \( T_{eff} \) and log g. For each \( T_{eff} \)
and log g, we computed the Br\( \gamma  \) synthetic line profile using the 
SYNSPEC code developed by Hubeny (Hubeny \citealp{Hubeny1};
Hubeny \& Lanz \citealp{Hubeny2}). These photospheric line profiles are then broadened
by solid rotation and can be further absorbed by the envelope volume projected
on the stellar disk. The SIMECA code is also able to produce theoretical intensity
maps of the circumstellar envelope in this line and in the continuum at different
wavelengths. These maps can be directly compared to milli-arcsecond interferometric 
measurements, such as those obtained from the VLTI interferometer.

\noindent For the m1 parameter, which describes the 
variation of the mass flux ($\phi$) from the pole to the equator, we used m1=20 (see
Stee \citealp{stee4} for more details) according to:

\begin{equation}
\label{flux}
\phi (\theta )=\phi _{pole}+[(\phi _{eq.}-\phi _{pole})sin^{m1}(\theta )]
\end{equation}

\noindent where \( \theta  \) is the stellar colatitude. This m1 value corresponds to an opening angle of 30$\degr$ and thus to a relatively flat disk, as can be seen in the lower row of 
Figs. \ref{intensitymaps} and \ref{intensitymaps_flux}. 
Since the contribution of the polar stellar wind is very faint and the terminal velocity in the dense equatorial region
is only 100 kms$^{-1}$, the disk is nearly Keplerian with a rotational velocity of  350 kms$^{-1}$ at the stellar photosphere. 
It produces a typical double-pics emission line profile with intensity maps that are strongly dependent on the observing wavelength as can be seen in Fig. \ref{visi_baseline_radius}. The basic parameters used to simulate the Be circumstellar disk in the following are given in Table 1.

\section{The ring scenario}
In order to simulate the excavation of the disk, we computed a set of static models 
where the density is forced to be at a very low value (nearly zero) from the stellar photosphere
to a given stellar radius, simulating the evolution of the gradual erosion of the entire disk as can be seen 
in Fig. \ref{intensitymaps}.   Each map is computed  at a wavelength centered on the Br$\gamma$ line 
within a spectral bandwidth of 5 $\AA$ and are plotted for 3 different inclination angles.

The shape of the intensity maps from Fig. \ref{intensitymaps} ranges from a pure ring-like geometry for the pole-on case to
a ``bipolar" shape for the intermediate inclination angle (i=45$\degr$) and the equator-on case. The line profiles from Fig. 
\ref{lines_visibilities_ring} present a single peak for the pole-on case, whereas the increasing inclination angle produces a double-peaked
line profile with a larger peak separation for a larger inclination angle. The small absorption close to the center of the Br$\gamma$ line, on the red side of the
line for the edge-on case, is a numerical artifact. In this simulation the assumed central star radius of 6 R$ _{\sun }$ is placed at a 
distance of 100 pc, and the Br$\gamma$ circumstellar disk is resolved for baselines as small as 30 meters (for a baseline parallel  to the
equatorial disk). Of course the disk is not resolved as well if the interferometer baseline is perpendicular to the equatorial disk as shown in Fig.
\ref{visi_para_perp}. In this latter case the interferometer is not able to ``see" the disk dissipation since its angular resolution perpendicular to
the equatorial flat disk is not sufficient and the visibilities are similar to those obtained for a uniform disk with a decreasing angular size.
The SED curve plotted in Fig. \ref{sed_radius} shows the decrease in the IR-excess as a function of the ring extension down to the value where
the continuum is only due to the central star (lower plain-line), which is assumed to radiate as a 22500 K blackbody.\\

\begin{figure*}[t]
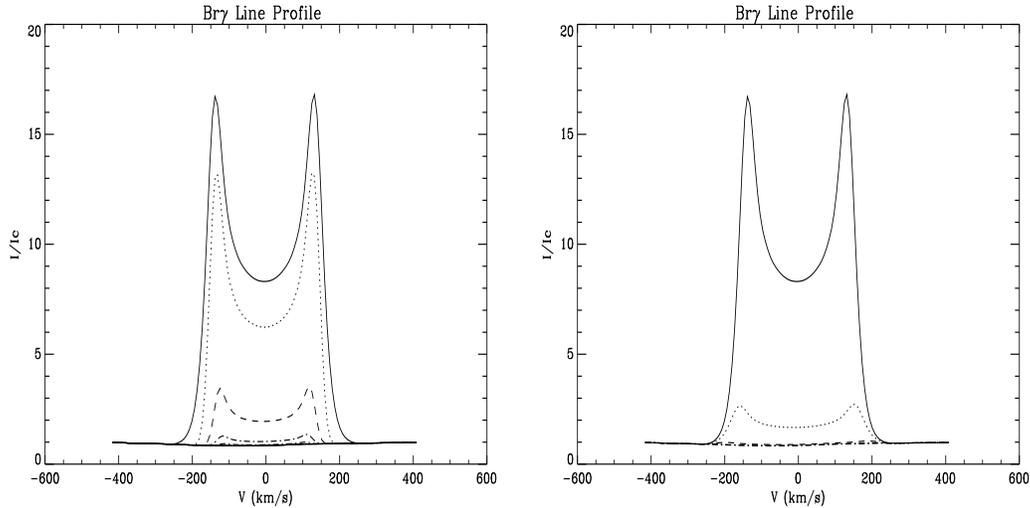

  \begin{center}
     \includegraphics[width=7cm,height=7cm]{4690F024.ps}
     \includegraphics[width=7cm,height=7cm]{4690F044.ps}
  \end{center}
  \caption{Br$\gamma$ line profiles from the ring (left) and mass flux scenarios computed for an inclination angle of 45$\degr$ between the star rotation axis and the observer's line of sight as a function of the disk dissipation.}
\label{compare_lines}
\end{figure*}

\begin{figure*}[t]
  \begin{center}
     \includegraphics[width=7cm,height=7cm]{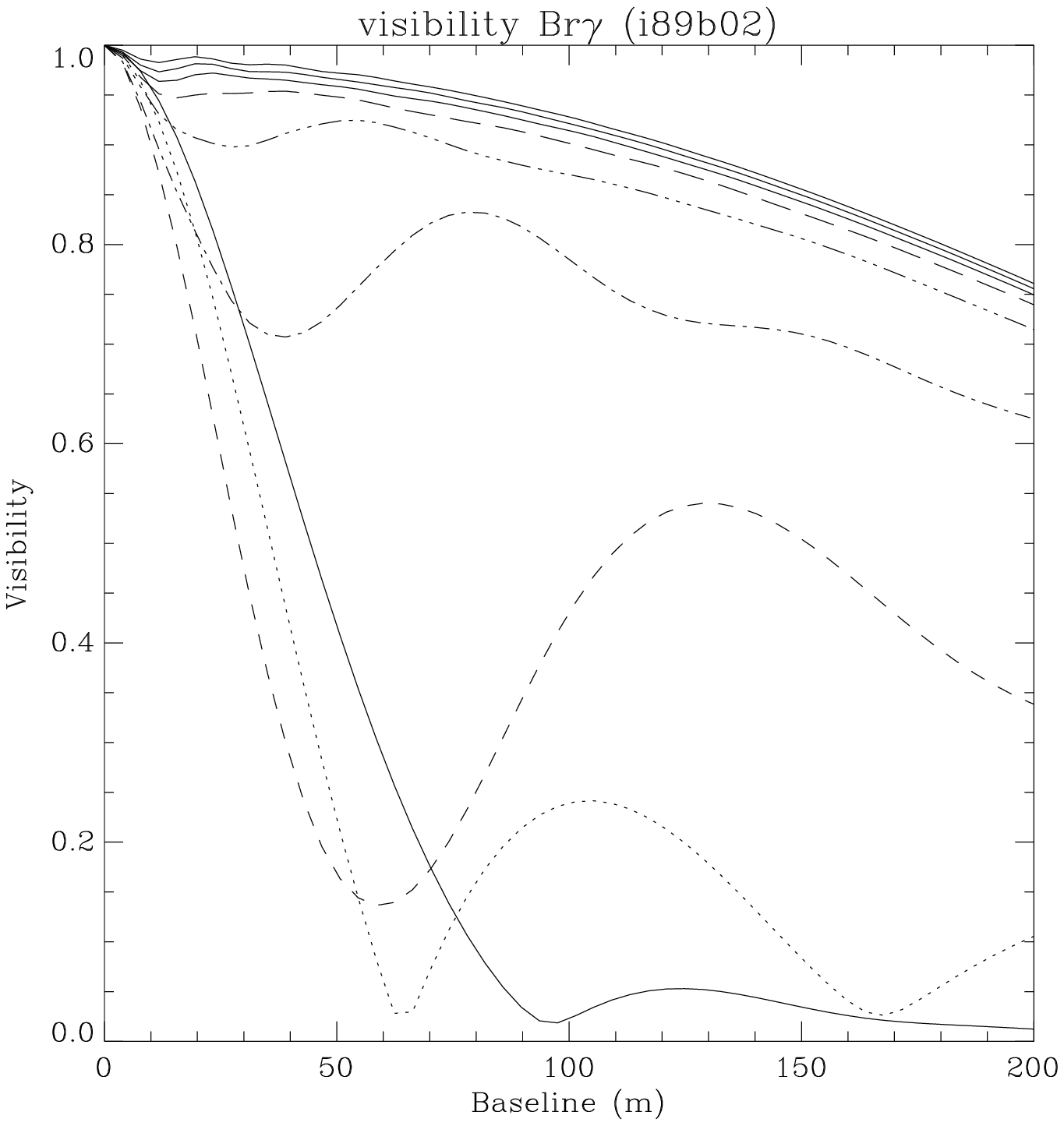}
     \includegraphics[width=7cm,height=7cm]{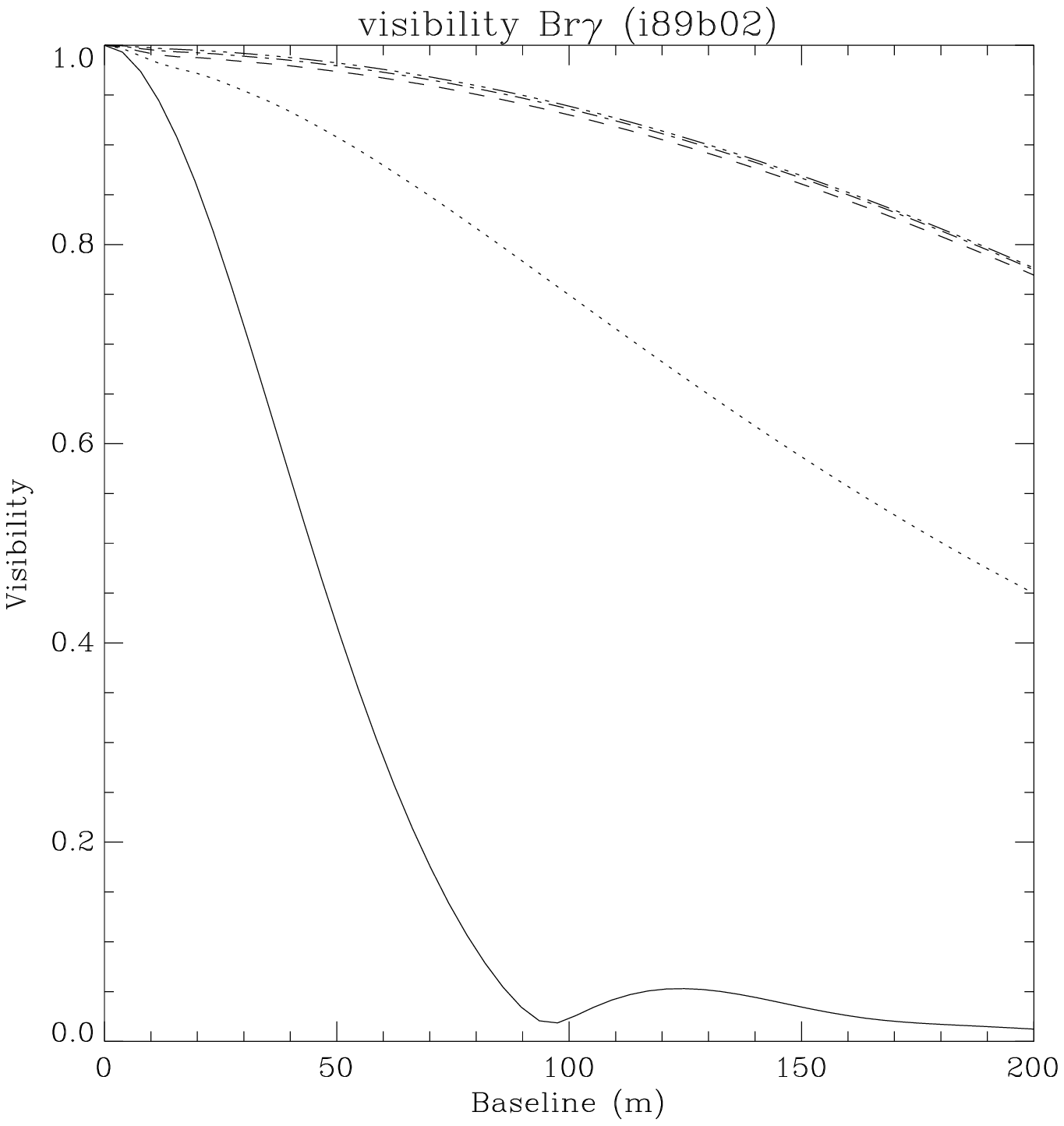}
  \end{center}
  \caption{Visibilities as a function of the baseline length for a baseline orientation along the equatorial disk as a function of the disk dissipation: following the ring model (left) and the decreasing mass flux scenario (right)}
\label{compare_visi}
\end{figure*}

\begin{table}
\centering 
\begin{tabular}{|c||c|}
\hline 
T\( _{eff} \)&
22500 K\\
\hline 
Radius&
6 R\( _{\sun } \)\\
\hline 
Inclination angle i&
0, 45 \& 90\ensuremath{}\\
\hline 
Photospheric density&
1.0 10\( ^{-12} \)g cm\( ^{-3} \)\\
\hline 
Photospheric expansion velocity&
0.11 km s\( ^{-1} \)\\
\hline 
Equatorial rotation velocity&
350 km s\( ^{-1} \)\\
\hline 
Equatorial terminal velocity&
100 km s\( ^{-1} \)\\
\hline 
Polar terminal velocity&
2000 km s\( ^{-1} \)\\
\hline 
Polar mass flux&
4.0 10\( ^{-12} \)M\( _{\sun } \) year\( ^{-1} \) sr\( ^{-1} \)\\
\hline 
m1&
20.0\\
\hline 
Mass of the disk&
5.06 10\( ^{-10} \)M\( _{\sun } \)\\
\hline 
Mass loss&
4.7 10\( ^{-10} \)M\( _{\sun } \) year\( ^{-1} \)\\
\hline 
\end{tabular}
\caption{Parameters and results used in this paper to simulate a Be circumstellar disk}
\end{table}

\section{The vanishing mass flux scenario}
In this scenario, the mass loss maybe slowly decreases due to a decrease in the radiative
force through an opacity change at the base of the photosphere. In this scenario the mass ejected 
in the circumstellar disk is becoming smaller and smaller, and this decreasing
brightness contrast between the central star  and its surrounding produces a fading of the emitting line profiles
(Fig. \ref{lines_visibilities_flux}), as well as of the IR excess (Fig. \ref{sed_flux}). The shape of the Br$\gamma$ lines
is very similar to those obtained for the ring scenario, since the envelope kinematics remains the same for both models.
Again, the small absorption close to the center of the Br$\gamma$ line, on the red side of the line for the edge-on case, 
is a numerical artifact. Reducing the global mass loss rate by a factor two, namely from  4.7 10$^{-10}$ to 2.3 10$^{-10}$ 
M$ _{\sun }$ year$^{-1}$ has a strong impact on the intensity of the line profiles, which goes from 95 to 15 times the continuum value
for the pole-on case and from 13 to 2 for the equator-on model. This also produces a drastic visibility change, and the disk is
fully resolved whatever the inclination angle is with baselines of 70 meters for the pole-on case and 100 meters for the equator-on model for
a mass loss rate of 4.7 10$^{-10}$ M$ _{\sun }$ year$^{-1}$. For the case of the 2.3 10$^{-10}$ M$ _{\sun }$ year$^{-1}$ mass-loss rate, the
visibilities never went to zero for baselines between 0 and 200 meters and the disk appears less resolved for larger inclination angles, i.e. from the 
pole to the equator. All the visibilities plotted in Fig. \ref{lines_visibilities_flux} are given for an interferometer baseline along the equatorial disk, i.e. 
horizontal with respect to the intensity maps from Fig. \ref{intensitymaps_flux}.

  \begin{figure}
  \begin{center}
     \includegraphics[width=8cm,height=8cm]{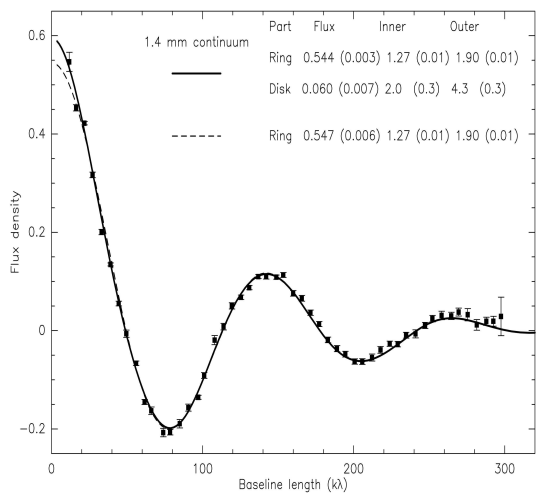}
  \end{center}
  \caption{GG Tauri real part of the 1.4 mm continuum visibility  as a function of the IRAM baseline length (dots with error bars)
  with a model (plain line) where 90\% of the flux comes from a ring and the rest from an extended disk from Fig. 2 in Guilloteau et al. \cite{guilloteau}. Note that this is the real part 
  of the visibility (which can be negative!) whereas in our simulation we plotted the visibility modulus that is always between 0 and 1. 
  The global shape of the visibility curve is the same as for our ring model from Fig. \ref{lines_visibilities_ring} seen pole-on.}
\label{ggtau}
\end{figure}

\section{Comparison of these two scenarios}
Since the aim of this paper was to see if there is a clear difference between the observable (photometric, spectrometric, and
interferometric) produced from these two scenarios, we may first compare the SEDs. It is clear from Figs. \ref{sed_radius} and \ref{sed_flux} that we cannot separate the scenarios: since the envelope is optically, thin there is no difference
with respect to the inclination angle; and the continuum produced from a given hydrogen volume is more or less the same, 
whereas it is distributed in a star-connected envelope or in a circumstellar ring. The differences from the line profiles are more
subtle as can be seen in Fig. \ref{compare_lines}.

In the case of the ring dissipation, the disappearance of the high velocity tails in the emission line is obvious and was already observed by
Rivinius et a. \cite{rivinus2} for 28 Cyg. This effect is due to the dense Keplerian equatorial disk 
with a monotonically decreasing rotational velocity. In this case, the emission from the innermost disk region will appear (or disappear) at
the outermost line wings. In this paper, we reach the same conclusion as in Rivinus et al. \cite{rivinus2} that the variability in the emission base
width is due to density variations in the inner disk. The position of the two line peaks remain mostly unchanged since they originate in the
outermost regions of the disk, which have the largest surface area in a given velocity bin. On the other hand, for the other scenario, the
diminution of the mass flux produces a global decrease in the line intensity without any noticeable variation in the base line width.
Since the line emission is proportional to the volume of the emitting gas, the line intensity decreases more rapidly in this
 scenario compared to the previous one, as already mentioned in the previous section.\\

 The  largest discrepancies come from the visibility curves. From Fig. \ref{compare_visi} we can see that
 the plain line that corresponds to the same starting model, i.e. an extended disk connected to the star surface,
 shows a fully resolved object with a visibility close to zero for the 100 meters baseline and a very
 small second lobe of about 5 \%. Starting with the same model, the differences are straightforward between the model with a ring at r=10 R$_*$ 
 and a mass-loss rate of 2.3 10$^{-10}$ M$ _{\sun }$ year$^{-1}$, i.e. a mass-loss decrease by a factor 2 compared to the previous 
 model. The visibility is resolved for the ring model for a short 65 meters baseline due to the fact that 
 there is less circumstellar matter close to the central star that is also unresolved by the interferometer. Thus, the flux ratio between the unresolved
 star+inner-envelope and the extended resolved outer-disk decreases and it is easier to ``see" this circumstellar outer-matter. This effect also
 increases the value of the first minimum of the visibility curve which goes from 1\% to 3\% and reaches  up to 15 \% and 70 \% for the r=20 R$_*$ and r=30 R$_*$ rings, respectively. The second lobe increases as a function
 of the ring radius up to the case where the flux coming from the ring is small enough to become negligible compared to the
 stellar flux. In this case the curves converge to a nearly spherical and unresolved object (the central star) that starts 
 to be resolved for baselines on the order of 100-200 meters. The shape of the second lobes for the ring model occurs mainly due because we simulate 
 a very sharp inner edge to the disk, i.e. we artificially set the density to zero in the inner ring region. The ``true" situation may be less 
 contrasted since the inner edge might be smoother, thereby making a less marked second lobe of the visibility. The visibilities for the
 second scenario are more regular with a global tendency to show a ``smooth" object, i.e. a more or less Gaussian  brightness 
 distribution and a less and less resolved object. The final shape of the visibilities is similar in both scenarios since the
 situation converges to the same visibility as the nearly unresolved central star. 
 
 \section{Conclusions}
 It is clear from this study that, in order to definitely favor the ring scenario of disk dissipation instead of the
 mass-loss vanishing, we must follow both spectroscopically and interferometrically a Be star that is showing a
 decrease in its IR excess, as well as in its emission line intensities.
 
 A clear signature of the disk dissipation following the ring scenario could be:\\
 -the disappearance of the high velocity tails in the emission line and a nearly constant peak separation;\\
 - visibilities with an increasing second lobe, an increase in the value of the first zero, and, assuming an unresolved
 central star, a first zero at shorter baselines as a function of the disk excavation.\\
 
 Thanks to the VLTI and its AMBER instrument, which combine high spatial resolution (2 mas for the 200m baseline) 
 and a spectral resolution of 10000, the ``direct" observation of a varying
 inner disk radius, certainly correlated to circumstellar outbursts, will be soon accessible. This kind of observation has already be done
by Guilloteau et al. \cite{guilloteau}  for GG Tau using  the IRAM interferometer.  We can see in Fig. \ref{ggtau} that the
 real part of the 1.4 mm continuum visibility  as a function of the IRAM baseline length (dots with error bars) agrees
 very well with a model (plain line) where 90\% of the flux is coming from a circumbinary ring and the rest from an extended disk. 
 The global shape of the visibility curve is globally the same as for our pole-on ring models from Fig. \ref{lines_visibilities_ring} 
 even if we have plotted visibilities modulus (between 0 and 1) in our modeling whereas this is the real part of the visibility,  which can be negative, that is plotted in Guilloteau et al. \citealp{guilloteau}.


\begin{acknowledgements}
We thank Anne Dutrey and St\'ephane Guilloteau for providing and allowing us to use their Fig. 2 in this paper. We also thank Damien Mattei for his help in developing the SIMECA code, as well as David Chapeau, Jean-Michel Clausse and Pascal Marteau for their support in the computer and network settings. J.R.R. Tolkien is also acknowledged for helping us to find the right title for this paper.
\end{acknowledgements}

\end{document}